\documentclass[final,3p,times]{elsarticle}

\usepackage{amssymb}
\usepackage{amsmath}
\usepackage{graphicx}
\usepackage{xcolor}
\usepackage{accents}
\usepackage{array}
\usepackage{color}
\usepackage{tikz}

\definecolor{myblue}{rgb}{0.12156862745098039, 0.4666666666666667, 0.7058823529411765}

\definecolor{myorange}{rgb}{1.0, 0.4980392156862745, 0.054901960784313725}

\definecolor{mygrey}{rgb}{0.7529411764705882, 0.7529411764705882, 0.7529411764705882}

\newcommand{\greydashedline}{\raisebox{2pt}{\tikz{\draw[-,mygrey,densely dashed,line width = 1.75pt](0,0) -- (5mm,0);}}}

\newcommand{\blueline}{\raisebox{2pt}{\tikz{\draw[-,myblue,solid,line width = 1.75pt](0,0) -- (5mm,0);}}}

\newcommand{\orangeline}{\raisebox{2pt}{\tikz{\draw[-,myorange,solid,line width = 1.75pt](0,0) -- (5mm,0);}}}

\newcommand{\thinblueline}{\raisebox{2pt}{\tikz{\draw[-,myblue,solid,line width = 1.25pt](0,0) -- (5mm,0);}}}

\newcommand{\thinorangeline}{\raisebox{2pt}{\tikz{\draw[-,myorange,solid,line width = 1.25pt](0,0) -- (5mm,0);}}}

\newcommand{\bluecircle}{\raisebox{0.5pt}{\tikz{\node[fill,scale=0.4,circle,fill=myblue](){};}}}

\newcommand{\orangecircle}{\raisebox{0.5pt}{\tikz{\node[fill,scale=0.4,circle,fill=myorange](){};}}}

\newcommand{\orangedashedline}{\raisebox{2pt}{\tikz{\draw[-,myorange,densely dashed,line width = 1.75pt](0,0) -- (5mm,0);}}}

\journal{ArXiv}

\begin{document}

\begin{frontmatter}




\title{Quantitative analysis of the kinematics and induced aerodynamic loading of individual vortices in vortex-dominated flows: a computation and data-driven approach}








\author{Karthik Menon\fnref{label1}}
\author{Rajat Mittal\corref{cor0}\fnref{label1}}
\cortext[cor0]{Corresponding author}
\ead{mittal@jhu.edu}

\address[label1]{Department of Mechanical Engineering, Johns Hopkins University, Baltimore, MD 21218, USA}

\begin{abstract}
A physics-based data-driven computational framework for the quantitative analysis of vortex kinematics and vortex-induced loads in vortex-dominated problems is presented. Such flows are characterized by the dominant influence of a small number of vortex structures, but the complexity of these flows makes it difficult to conduct a quantitative analysis of this influence at the level of individual vortices. The method presented here combines machine learning-inspired clustering methods with a rigorous mathematical partitioning of aerodynamic loads to enable detailed quantitative analysis of vortex kinematics and vortex-induced aerodynamic loads. We demonstrate the utility of this approach by applying it to an ensemble of 165 distinct Navier-Stokes simulations of flow past a sinusoidally pitching airfoil. Insights enabled by the current methodology include the identification of a period-doubling route to chaos in this flow, and the precise quantification of the role that leading-edge vortices play in driving aeroelastic pitch oscillations. 
\end{abstract}


\end{frontmatter}


\section{Introduction}
\label{sec:intro}

The dynamical influence of vortex structures is key to a wide range of fluid-flow phenomena \citep{Peacock2013LagrangianFlows,Ellington1996Leading-edgeFlight,Eaton1994PreferentialTurbulence}. This is particularly true in vortex-dominated flows, where coherent vortex structures and their interactions exert a dominant influence on the force/moment production on immersed surfaces. In the case of fluid-structure interaction problems in particular, these vortex structures can drive the motion of immersed bodies, which in turn leads to the generation of additional vortices, and gives rise to complex non-linearities in the structural response. Such behaviour is relevant in a number of arenas including bluff-body oscillation, wing flutter, biological propulsion, and physiological flows \citep{Williamson1996VortexWake,Triantafyllou2000HydrodynamicsSwimming,Wang2005DissectingFlight,Mittal2018MattersHeart,Eldredge2019}.

\begin{figure}
  \centerline{\includegraphics[scale=1.0]{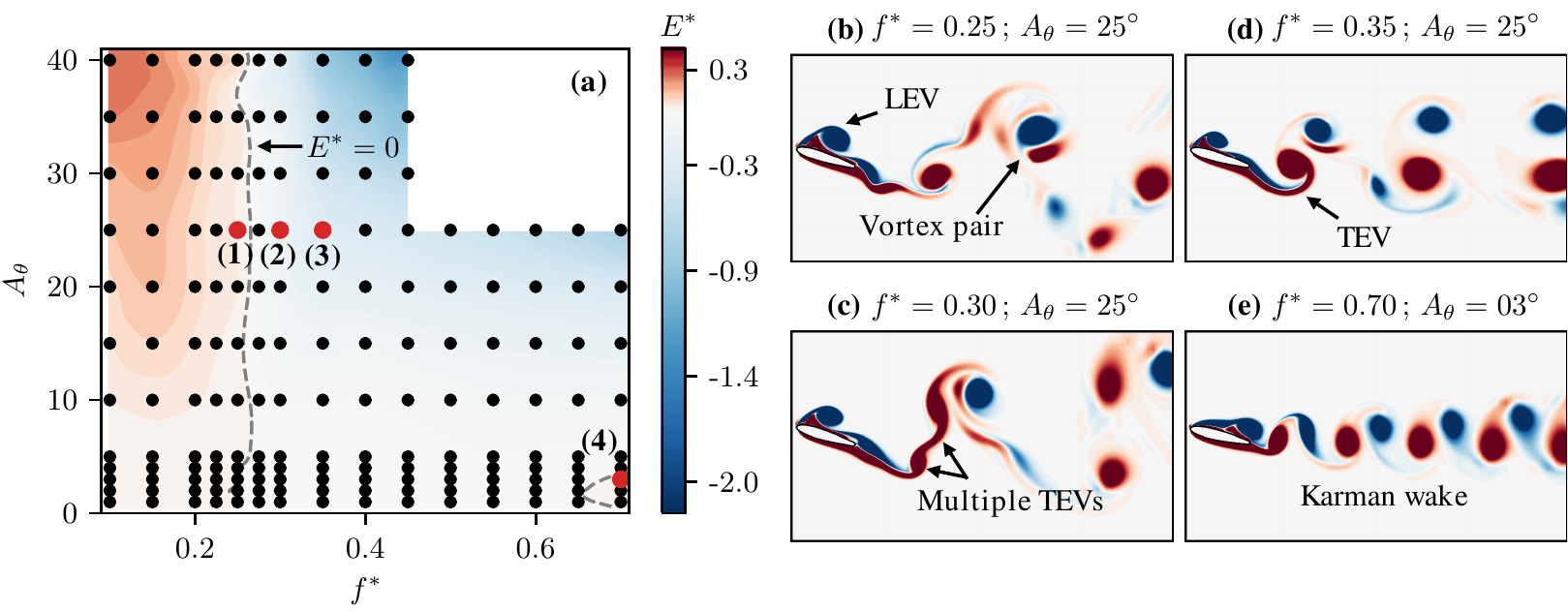}}
  \caption{Data set showing 165 two-dimensional Navier-Stokes simulations of an airfoil pitching sinusoidally about its mid-chord. The oscillation amplitude is $A_{\theta}$ and dimensionless frequency is $f^* = fC/U_{\infty}$ (where $C$ and $U_{\infty}$ are chord-length and free-stream velocity). Frequency and amplitude for each case in the data set is shown as circles in (a). Coloured contours in (a) show mean energy extracted by the oscillating airfoil over a cycle, $E^*$, defined in equation \ref{eq:energy}. The contour corresponding to $E^*=0$ is shown as dashed curve in (a). Figures (b)-(c) show snapshots of the vorticity field for select cases in this data set, with $f^*$ and $A_{\theta}$ for each case specified.}
\label{fig:vortex_intro}
\end{figure}

A prototypical problem that manifests much of the complexity associated with such flows -- vortex dominated behavior and complex vortex interactions -- is the flow around a pitching airfoil. To illustrate this, figure \ref{fig:vortex_intro}(a) shows a data set consisting of 165 two-dimensional Navier-Stokes simulations of flow past a sinusoidally pitching airfoil. This ensemble of cases represents a parameter sweep through the pitching frequency, $f^*$, and amplitude, $A_\theta$ (both defined in the caption of figure \ref{fig:vortex_intro}). The complex vortex dynamics inherent in such a problem is highlighted by the snapshots of the vorticity field shown for some select cases in figures \ref{fig:vortex_intro}(b)-(e). These snapshots show that this problem is characterized by a variety of vortex patterns that are quite sensitive to changes in oscillation kinematics. The snapshots in figures \ref{fig:vortex_intro}(b)-(d), corresponding to the same oscillation amplitude and relatively small differences in oscillation frequency, all show the growth of a strong leading-edge vortex (LEV), along with several other distinct, interacting vortices. However, we see that the phase of the LEV-growth is slightly different in each case, although the snapshots are at the same phase in the oscillation. 

One way to assess the overall aeroelastic interaction of the flow with the airfoil is to evaluate the energy that could potentially be extracted by the airfoil from the surrounding flow as a function of oscillation kinematics. The contours in figure \ref{fig:vortex_intro}(a) show this energy extraction, which is defined as, 
\begin{equation}
E^* = \int^{1/f^*}_{0} \, C_M \, \dot{\theta} \, dt
 \label{eq:energy}
\end{equation}
where $C_M$ is coefficient of moment about mid-chord, and $\dot{\theta}$ is the angular velocity (see ref. \cite{Menon2019} for details). This energy is known to determine the flow-induced oscillation response as well as energy-harvesting potential of the airfoil \citep{Menon2019,Zhu2020NonlinearWing}. We see in figure \ref{fig:vortex_intro}(a) that kinematic states with positive as well as negative energy extraction are possible, which is primarily dictated by the phase difference between the dominant aerodynamic loading mechanisms and the oscillation kinematics \citep{Menon2019,Menon2020DynamicAirfoil}. As evidenced by these energy contours, the slight variation in oscillation kinematics for the cases shown in figures \ref{fig:vortex_intro}(b)-(d), which results in subtle changes in the phase of the LEV as well as different vortex interactions,  has consequences for the sign of energy extraction and therefore the dynamics of flow-induced motion. Additionally, all these cases exhibit very different vortex interactions close to the leading and trailing-edges, and this can influence the force production, propulsion, and fluid-structure interaction \citep{Anderson1998OscillatingEfficiency,Zhu2002Three-dimensionalSwimming,Martin-Alcantara2015VortexAttack} associated with such configurations.

Thus we see that these flows are generally characterized by several interacting vortex structures. The kinematics and the dynamical influence of these vortices on an immersed/control surface are governed by their inception, phase, location, as well as their interactions with each other and the immersed surface. While prior studies have highlighted the dynamical importance of specific vortex structures such as the LEV \citep{Eldredge2019}, as well as distinct patterns in vortex shedding \citep{Williamson1996VortexWake}, they have largely been limited to \emph{qualitatively} correlating the occurrence and evolution of these structures to the observed dynamics of the problem. The question of precisely \emph{quantifying} their dynamical influence, in terms of force and moment production on an immersed body for instance, has not been adequately addressed. Moreover, a systematic and rigorous way to partition the influence of the various vortex structures generally present in such problems, and to identify those that are the most dynamically important to that specific problem, do not currently exist.

The analysis of these general vortex-dominated flows, which are usually characterized by several distinct, interacting vortices therefore requires two key elements: (1) the isolation, tracking and segmentation (i.e. determination of size and shape) of multiple individual vortex structures that are generated in the unsteady flow; (2) the rigorous quantification of the aerodynamic forces and moments induced on an immersed body by each of these vortex structures. In this work, we propose a computational framework to perform such an analysis of high-dimensional, time-resolved flow-fields at the level of individual vortex structures. In particular, we combine a rigorous force and moment partitioning method (FMPM) which enables the precise estimation of the aerodynamic loads induced by individual vortices, with data-driven techniques that facilitate the efficient use of this method in complex vortex-dominated flows. The result of this combined physics-based data-driven approach is a versatile and largely automated framework that can decode the vortex kinematics and dynamics of such problems by isolating each vortex structure in the flow, and evaluating its dynamical effect on an immersed body through its entire spatio-temporal evolution.

A central piece in this analysis framework is a mathematically rigorous method for partitioning fluid dynamic forces and moments on an immersed body into contributions from individual vortices, as well as other viscous and inviscid forcing mechanisms. The method used here, which is based on an exact analytical formulation derived from the Navier-Stokes equations, follows from the work of Quartappelle and Napolitano \cite{Quartappelle1982ForceFlows}, with extensions by Chang \cite{Chang1992PotentialFlow} and Zhang et al. \cite{Zhang2015CentripetalInsects} for the partitioning of flow-induced forces on an immersed body into physically relevant mechanisms. Here we extend the formulation that has been applied to flow-induced forces, to also include the partitioning of flow-induced \emph{moments}. This is particularly relevant for problems with rotational/pitch degrees-of-freedom. Furthermore, the specific formulation developed in this work results in force/moment components that have clear physical interpretations, thereby allowing us to relate, using first principles, the mechanisms behind force/moment generation in incompressible flows to the local kinematics of the flow. We note that the partitioning of flow-induced pressure forces used here is not unique, and there have been other mathematically rigorous force partitioning formulations \cite{Wu1981TheoryFlows,Noca1999ADerivatives}, as well as extensions to the formulation used here \citep{Howe1995OnNumbers,Protas2000AnFlows,Pan2002AFlow,Magnaudet2011ANumber}.

In the context of analyzing force production in vortex dominated flows, these partitioning methods have proven to be very useful in delineating the overall contribution of vortex-induced effects, as well as other physically relevant forcing mechanisms, in various unsteady aerodynamics and fluid structure interaction problems \citep{Wu2007IntegralStructures,Zhang2015CentripetalInsects,Martin-Alcantara2015VortexAttack,Moriche2018OnNumber,Menon2020OnPartitioning}. While there have also been other efforts to quantify the force production from vortex-induced mechanisms, these have primarily used inviscid vortex models and vortex impulse formulations \cite{Wang2013Low-orderFormation,Graham2017AnFlow}, or simple inviscid theory to separate added-mass from so-called ``vortex-flow force'' \citep{Lighthill1986FundamentalsStructures,Govardhan2000,Carberry2005ControlledModes}. However, conceptual validity of these latter methods has been questioned \citep{Sarpkaya2001OnMorison}, and the extension of these methods to general viscous flows is unclear. Further, these prior efforts have mostly focused on analyzing the \emph{global} effect of vorticity in the flow. Estimations of force/moment production due to \emph{local} vortical regions have thus far been limited to determining vortex-induced forces within static spatial volumes of the flow-field \cite{Wu2007IntegralStructures,Menon2020OnPartitioning}, rather than individual vortices. This is primarily due to the computational complexity associated with accurately isolating and tracking several interacting vortical regions that are evolving in highly complicated, time-varying flow fields. Therefore, as mentioned earlier, a critical aspect in deploying these rigorous force/moment estimation methods for the analysis of highly dynamic vortex-dominated flow-fields is a systematic way to individually isolate and track these vortical regions.

Here we leverage data-driven clustering techniques to perform this task of isolating and tracking several vortices in high-dimensional, unsteady flow-fields. These techniques are a class of unsupervised statistical inference methods \citep{Jain1999DataReview,han2011data} which attempt to find clusters of data that share similar characteristics within a large set of data. In previous studies, this has been used in identifying coherent structures in a flow-field, by clustering regions that share similar dynamic or Lagrangian behaviour \citep{Nair2015Network-theoreticDynamics,Froyland2015AData,Ser-Giacomi2015FlowTransport,Hadjighasem2016Spectral-clusteringDetection,Schlueter-Kuck2017CoherentTheory,Padberg-Gehle2017Network-basedMixing}. In the context of this work, clustering provides an automated way to isolate an arbitrary number of spatial regions corresponding to individual vortex structures from high-dimensional flow-fields. Furthermore, this data-driven approach also facilitates the spatio-temporal tracking of these vortex structures, as well as the grouping of vortices in distinct  categories (such as LEVs or TEVs) based on various attributes. The automated isolation of these distinct vortical regions therefore provides an efficient way to employ the aforementioned force and moment partitioning methods in precisely quantifying their contributions to force/moment production. Furthermore, the spatio-temporal tracking of each of these structures also allows us to relate their evolution and interactions to the dynamics of the problem, which has been shown to be very insightful in past work \citep{Huang2015DetectionStructures,Rockwood2017DetectingStructures,Rockwood2018TrackingFlows}. Hence this combined physics-based and data driven approach provides an automated and rigorous method to analyze vortex kinematics as well as the force and moment production in complex vortex dominated flows. 

While the framework described above allows the detailed dynamical analysis of a single flow, the automated nature of the method makes it particularly well suited for examining a large ensemble of flows. We demonstrate this by applying our method to the data set generated from the 165 pitching airfoil simulations represented in figure \ref{fig:vortex_intro}. In the current study, application of the method to this large ensemble is preceded by a data-driven reduction in the ``rank'' of the data set via identification of distinct vortex-dynamic regimes in this flow. This identification of distinct regimes has similarities to previous studies on the wake of oscillating cylinders \citep{Zdravkovich1981ReviewShedding,Ongoren1988FlowWake,Williamson1988VortexCylinder,Williamson1996VortexWake,Williamson2004Vortex-InducedVibrations} as well as in biomimetic propulsion \citep{Koochesfahani1989VorticalAirfoil,Drucker1999LocomotorVelocimetry,Triantafyllou2004ReviewFoils}. However, the visual identification employed in these past studies is impractical for more complex flows; a fact that has motivated data-driven approaches to this problem  \citep{Hemati2016LearningData,Colvert2018ClassifyingNetworks,Wang2019DetectingSensors}. However, these prior data-driven efforts have focused on idealized wakes, using point vortices for example, and have mostly assumed a-priori knowledge of the possible wake patterns in order to classify observed wakes into these known categories. Here we demonstrate a more robust method to identify distinct flow regimes from high-dimensional, time-resolved flow-fields by using dimensionality reduction and clustering techniques to identify patterns in large ensembles of these flow-fields. This approach presented here is robust and also obviates the need for a-priori definition of wake-patterns.

Hence, the framework developed here allows the analysis of a large ensemble of flows at the resolution of individual vortices, and includes a quantification of the vortex kinematics as well as and their dynamical influence. In the following sections, we first describe the methods developed to enable this analysis, i.e. the force and moment partitioning method, the vortex isolation, tracking and dynamical analysis methodology, and lastly the vortex-regime identification procedure. This description of the methods is then followed by an application of these methods to the pitching airfoil data set mentioned above, followed by concluding remarks. 

\section{Force and moment partitioning method}
\label{sec:fpm}
\newcommand{\fpmkin}{C^{\kappa}_{F_i}}
\newcommand{\fpmvif}{C^{\omega}_{F_i}}
\newcommand{\fpmshr}{C^{\sigma}_{F_i}}
\newcommand{\fpmpot}{C^{\Phi}_{F_i}}
\newcommand{\fpmext}{C^{\Sigma}_{F_i}}

\newcommand{\mpmkin}{C^{\kappa}_{M_k}}
\newcommand{\mpmvif}{C^{\omega}_{M_k}}
\newcommand{\mpmshr}{C^{\sigma}_{M_k}}
\newcommand{\mpmpot}{C^{\Phi}_{M_k}}
\newcommand{\mpmext}{C^{\Sigma}_{M_k}}
\newcommand{\vol}{{V_f}}

\begin{figure}
  \centerline{\includegraphics[scale=1.0]{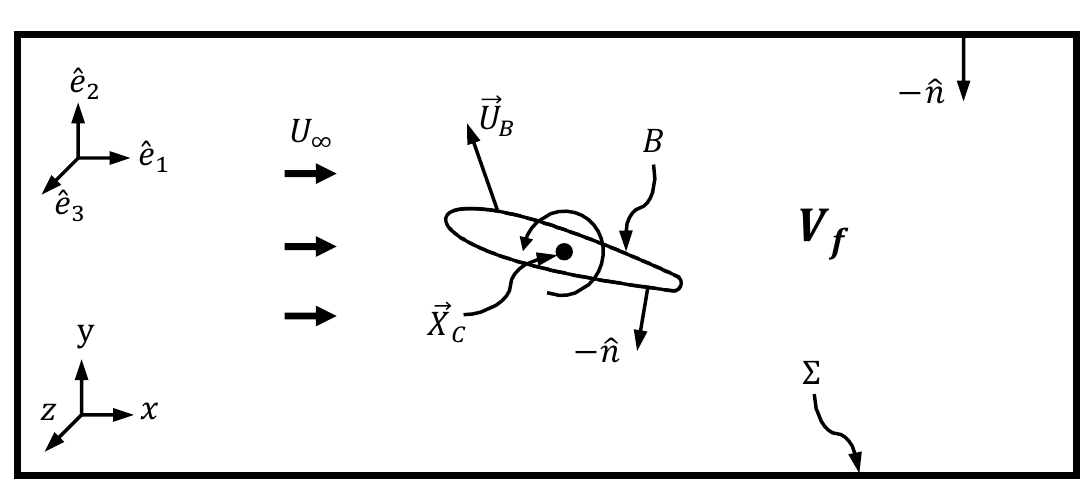}}
  \caption{Schematic of the problem setup for the force and moment partitioning method, along with relevant symbols.}
\label{fig:fpm_schematic}
\end{figure}
We first provide an overview of the force and moment partitioning methods (FMPM) used in this work, since this method drives much of the data-driven methodology described in the current paper. The partitioning is described with reference to the configuration shown schematically in figure \ref{fig:fpm_schematic}, where we have a body with its surface represented as $B$, immersed in a fluid domain of volume $V_f$. The unit normal vector $\hat{n}$ points into the surface $B$ at every point along it, and the fluid flow around this body induces aerodynamic loads on its surface. The aerodynamic force and moment induced by the pressure ($p$) on the immersed body are given by,
\begin{equation}
 \vec{F}_p =  \int_{B} p \, \hat{n} \, dS  \quad \text{and} \quad
 \vec{M}_p = \int_{B} p \left[ \left(\vec{X}-\vec{X}_c \right)\times\hat{n} \right] \, dS, 
 \label{eq:surface_integral}
\end{equation} 
respectively, where $\vec{X}$ is the position vector of a point on the surface of the immersed body and $\vec{X}_c$ is the point about which the moment is calculated. These forces and moments are usually computed in a simulation by directly evaluating these integrals on the surface of the immersed body. However, the induced pressure on the surface incorporates the effect of various physically distinct mechanisms -- such as multiple vortices in its vicinity, shear/boundary layers, as well as added-mass and viscous diffusion effects; and it is not immediately obvious how the individual contributions of each of these mechanisms can be partitioned out from the total pressure induced force and moment. 

The method discussed here is based on the work of Quartappelle and Napolitano \cite{Quartappelle1982ForceFlows}, who showed that the pressure-induced forces and moments on an immersed body can be written in terms of velocity-field gradients by projecting the Navier-Stokes equations onto the gradient of an appropriately constructed auxiliary potential field. The physical relevance of the terms in this projection-based approach was later recognized by Chang \cite{Chang1992PotentialFlow} and Zhang et al. \cite{Zhang2015CentripetalInsects}. The current partitioning is based on these prior formulations, but the specific form derived and used here is different due to our emphasis on vortex-induced pressure. In this section we provide an overview of the key aspects of this method and a detailed derivation of the method is shown in \ref{app:fpm_derivation} and ref. \citep{Menon2020OnPartitioning}.


The force and moment partitioning is presented in dimensionless form, where the force/moment coefficients, as well as velocity-field quantities, are non-dimensionalized using the velocity scale $U_\infty$, and length scale $L$. Hence the force coefficient on the immersed body in the $i$-direction is given by $C_{F_i} = F_{i}/(\frac{1}{2} \rho U^2_\infty L)$, where $F_{i}$ is the dimensional form of this total force. Similarly, the moment coefficient in the $k$-direction is given by $C_{M_k} = M_{k}/(\frac{1}{2} \rho U^2_\infty L^2)$, where $M_{k}$ is the dimensional form of the total moment.

The method presented here partitions these force and moment coefficients into the following components:
\begin{equation}
    C_{F_i} = \fpmkin + \fpmvif + \fpmshr + \fpmpot + \fpmext \; \; \; ; \; \; \; C_{M_k} = \mpmkin + \mpmvif + \mpmshr + \mpmpot + \mpmext \; \; \; ; \; \; \;  i,k=1,2,3.
    \label{eq:fpm_mpm_decomp}
\end{equation}
Here the superscripts represent different physically identifiable components of the force/moment on the immersed body. The contribution associated with the kinematics of the body (i.e. acceleration-associated effects) are represented by the superscript $\kappa$, the vorticity-induced components by $\omega$, the effect of viscous shear and diffusion is denoted by $\sigma$,  effects due to the irrotational component of the flow are represented by $\Phi$, and the contribution from the flow and vorticity on the outer boundaries of the spatial domain is given by $\Sigma$.

In this work, we are primarily interested in evaluating the aerodynamic loads induced by vortex structures or, more generally, vorticity containing regions of the flow. We will therefore focus on the vorticity-induced components of the force ($\fpmvif$) and moment ($\mpmvif$) for the remainder of this discussion (the reader is referred to \ref{app:fpm_derivation} for the mathematical forms of the other terms). For the partitioning of forces in the $i$-direction and moments in the $k$-direction respectively, the vorticity-induced components due to volume $V_f$ of the fluid take the following form:
\begin{align}
    \fpmvif = - 2 \int_\vol Q \; \phi_{i} \; dV - \epsilon_{F_i}^{\Phi} \quad ; \quad \mpmvif = - 2 \int_\vol Q \; \psi_{k} \; dV - \epsilon_{M_k}^{\Phi} \label{eq:vif_combined_exact}
\end{align}
Here $\phi_i$ and $\psi_k$ are auxiliary potential fields defined over the domain of interest, which at every time-instance depend only on the instantaneous shape and position of the immersed body (defined by $B$ and $\hat{n}$). These fields are given by,
\begin{equation}
  {\nabla}^2 \phi_{i} = 0, \ \ \mathrm{ with } \ \
  \hat{n} \cdot \vec{\nabla} \phi_{i}=
    \begin{cases}
      n_i \;, \; \mathrm{on} \; B \\
      0 \; \;, \; \mathrm{on} \; \Sigma \\
    \end{cases};  
    \quad \quad
     {\nabla}^2 \psi_{k} = 0, \ \ \mathrm{ with } \ \
  \hat{n} \cdot \vec{\nabla} \psi_{k}=
    \begin{cases}
     \big[ (\vec{X}-\vec{X}_c)\times\hat{n} \big]\cdot \hat{e}_k \;, \; &\mathrm{on} \; B \\
      0 \; \;, \; &\mathrm{on} \; \Sigma \\
    \end{cases}
  \label{eq:fpm_scalar}
\end{equation}
where $n_i$ is the component of $\hat{n}$ in the $i$-direction and $\hat{e}_k$ is the Cartesian basis vector in the $k$-direction. In equation \ref{eq:vif_combined_exact}, $Q$ is second-invariant of the velocity-gradient tensor for the flow. This is usually defined as,
\begin{equation}
Q = \frac{1}{2}\left(||\boldsymbol{\Omega}||^2 - ||\boldsymbol{S}||^2 \right),
\label{eq:q-definition}
\end{equation}
where $\boldsymbol{\Omega}$ and $\boldsymbol{S}$ are the anti-symmetric and symmetric parts of the velocity-gradient tensor respectively. Positive values of $Q$ correspond to regions where rotation dominates over strain and vice versa. This rotation-dominance condition, $Q>0$, is commonly is used to detect vortices in a flow \citep{Hunt1988EddiesFlows}. 

Finally,  the terms $\epsilon_{F_i}^{\Phi}$ and $\epsilon_{M_k}^{\Phi}$ in equation \ref{eq:vif_combined_exact} explicitly partition the effects of purely irrotational flow from rotational flow-dependent contributions associated with $Q$ (see \ref{app:fpm_derivation} for details). Consequently, it can be shown that $\fpmvif$ and $\mpmvif$ are zero in the absence of vorticity (or rotational flow), and these terms hence correspond to the vorticity-induced component force and moment. Further, it can also be shown that $\epsilon_{F_i}^{\Phi} \approx 0$ and $\epsilon_{M_k}^{\Phi} \approx 0$ in most cases, and in particular for sufficiently large $V_f$ \cite{Zhang2015MechanismsInsects}. We therefore arrive at the following approximate form for the vorticity-induced force and moment, which will be used in the remainder of this work:
\begin{equation}
    \fpmvif \approx  - 2 \int_\vol Q \; \phi_{i} \; dV \quad ; \quad \mpmvif \approx  - 2 \int_\vol Q \; \psi_{k} \; dV
    \label{eq:vif_combined_approx}
\end{equation}

It is interesting to point out that the dependence of the vorticity-induced force and moment on $Q$ suggests a mechanism for the production of this force/moment component based on the local flow kinematics. Depending on the sign of $Q$, equation \ref{eq:vif_combined_approx} indicates that $\fpmvif$ and $\mpmvif$ are dictated by vorticity-induced strain and rotation in the flow. This is interesting because it shows that $Q$ is not just a metric describing local flow kinematics, but that it has a direct bearing on force production. This connection between $Q$ and pressure-induced force can be further reinforced by noting that $Q \equiv - (1/2) \vec{\nabla} \cdot \left( \vec{u} \cdot \vec{\nabla} \vec{u} \right)$. Therefore, the non-dimensional pressure Poisson equation for incompressible flow can be expressed as $\vec{\nabla}^2 p = 2Q$. Thus, $Q$ is in fact the source term in the pressure Poisson equation, and it is therefore not surprising that it also appears in the partitioned forces and moments.

\begin{figure}
  \centerline{\includegraphics[scale=1.0]{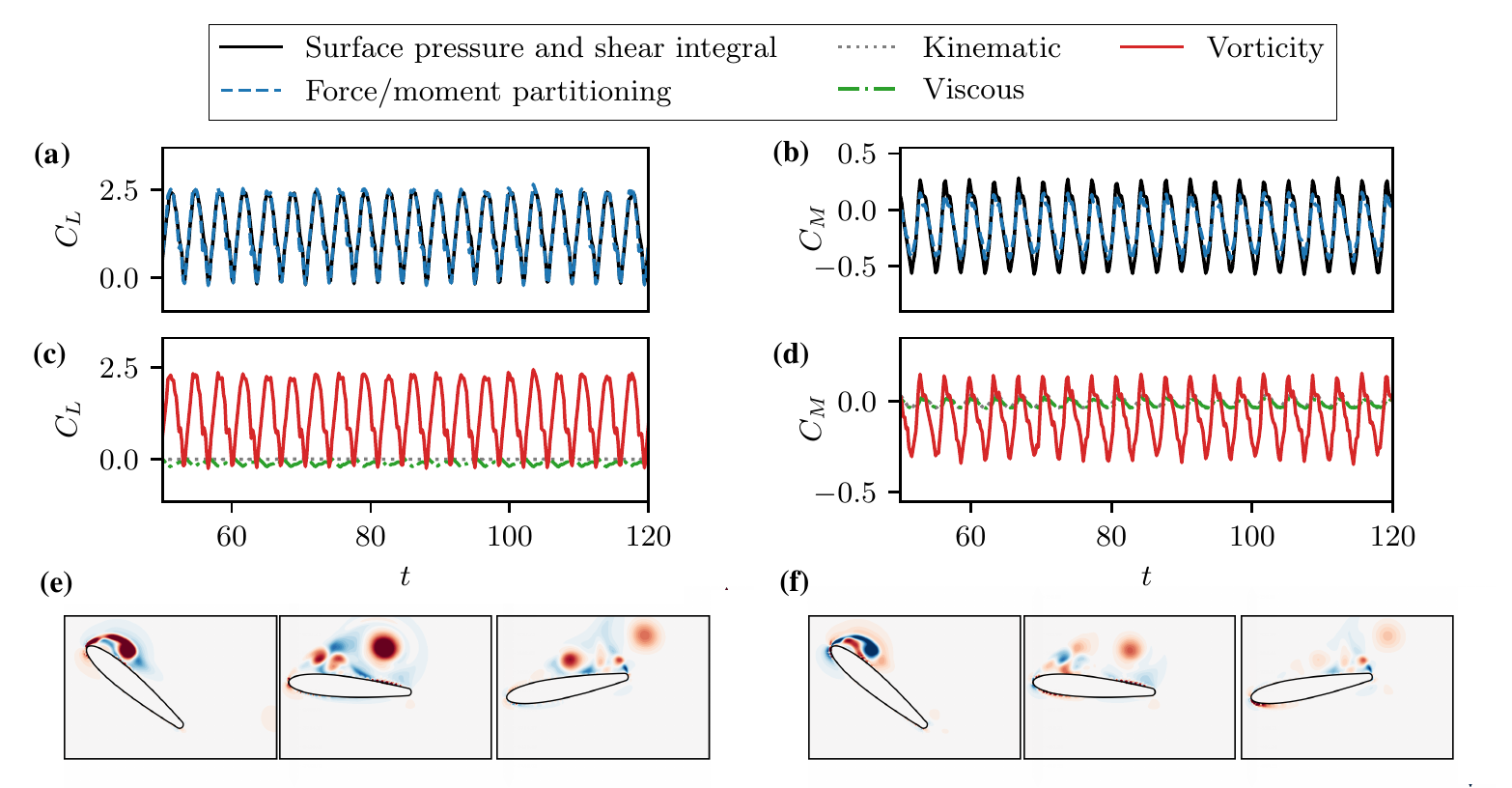}}
  \caption{Force and moment partitioning for a sample case of a sinusoidally pitching airfoil with dimensionless frequency $f^*=0.25$ and amplitude $A_\theta=25^{\circ}$. (a) Total lift coefficient ($C_L$) calculated using the conventional surface integral of pressure (equation \ref{eq:surface_integral}) and viscous shear, compared with sum of force partitioning terms, $C_L \approx C^{\kappa}_L + C_L^{\omega} + C_L^{\sigma}$; (b) Comparison of total moment coefficient ($C_M$) calculated using surface integral compared, with sum of moment partitioning terms, $C_M \approx C^{\kappa}_M + C_M^{\omega} + C_M^{\sigma}$; (c) Comparison of $\fpmkin$, $\fpmvif$ and $\fpmshr$; (d) Comparison of $\mpmkin$, $\mpmvif$ and $\mpmshr$; (e) Snapshots of vorticity-induced force distribution at three phases during a pitch-down cycle; (f) Snapshots of vorticity-induced moment distribution at three phases during a pitch-down cycle. }
\label{fig:fpm_mpm}
\end{figure}
We now present an example of the force and moment partitioning outlined above, for a sample case of a two-dimensional airfoil pitching  about mid-chord, at $Re = U_\infty C/\nu = 1000$ (where $C$ and $U_\infty$ are chord-length and free-stream velocity). The airfoil is forced to pitch sinusoidally with dimensionless frequency, $f^*=fC/U_\infty=0.25$, and amplitude $A_{\theta} = 25^{\circ}$. Figure \ref{fig:fpm_mpm}(a) shows a comparison of the total force coefficient in the vertical direction, i.e. the coefficient of lift $C_L$ (where $i=2$ and $C_{F_2} = C_L$), compared with the sum of the kinematic, vorticity-induced, and viscous components ($C^{\kappa}_L + C_L^{\omega} + C_L^{\sigma}$). Similarly, figure \ref{fig:fpm_mpm}(b) shows a comparison of the total pitching moment coefficient, $C_M$ (where $k=3$), with the sum of the terms ($C^{\kappa}_M + C_M^{\omega} + C_M^{\sigma}$). We see that the kinematic, vorticity-induced, and viscous components account for the bulk of the total force and moment production. In fact, it can be shown that irrotational and outer-boundary effects go to zero for sufficiently large domains (see ref. \cite{Zhang2015MechanismsInsects}). We also see from this comparison that, as expected in these vortex-dominated flows, the vorticity-induced component contributes a much larger lift and moment than the other two components. Lastly, figures \ref{fig:fpm_mpm}(e) and \ref{fig:fpm_mpm}(f) show snapshots of the vorticity-induced force and moment fields respectively, i.e. the integrands in equation \ref{eq:vif_combined_approx}, at three time-instances during the pitch-down motion of the airfoil. In both cases we see the strong influence of the leading-edge vortex (LEV), which we will quantify later in this work. In the context of the lift-force, the LEV induces a strong positive lift, whereas the moment due to the LEV changes sign as it crosses the pitch axis located at mid-chord. 

In addition to estimating the total aerodynamic loading due to these distinct mechanisms as in the example above, the volume-integral form of the terms in equation \ref{eq:vif_combined_approx} indicates that the contribution of any \emph{individual} vortex to the aerodynamic loading on an immersed body can be determined using equation \ref{eq:vif_combined_approx} by constructing the integral over the volume occupied by the vortex. The ability to construct integration volumes that isolate the effect of individual flow structures in highly dynamic vortex-dominated flows is itself a challenge, and the next section describes our approach to this task. We note that, in principle, the partitioned aerodynamic loads can be computed independent of the method used to generate these flow-fields, thereby allowing this method to be used with data from any simulation and potentially, even from an experiment. In this context, the force/moment partitioning method presented in this paper may be viewed as a data-driven method for the analysis of any flow-field with an immersed body. 

\section{Automated tracking of vortices and estimation of aerodynamic loads}
\label{sec:vortex_tracking}

While the force and moment partitioning method (FMPM) can determine the aerodynamic loading associated with any vorticity-containing region of the flow, complex vortex-dominated flows (such as in figures \ref{fig:vortex_intro} and \ref{fig:fpm_mpm}) typically contain multiple vortices which interact, deform as well as change their volume and location as they are advected with the flow. Thus, to effectively apply FMPM to such flows, appropriate methods are required to isolate, track, as well as determine the time-varying volumes occupied by each vortex in an automated manner. This is an important but non-trivial exercise, and a number of previous studies have developed tools for some aspects of this task, and demonstrated their utility in the analysis of vortex dominated flows \citep{Nair2015Network-theoreticDynamics,Huang2015DetectionStructures,Rockwood2017DetectingStructures,Rockwood2018TrackingFlows}. In this section, we describe a data-driven framework for automated tracking of vortices in relatively complex flows that is specifically suited to the application of FMPM. This procedure takes in time-resolved flow-field data and (1) isolates and tracks multiple individual vortex structures; (2) identifies and groups vortical structures that occur repeatedly (or periodically) in the given flow-field data; and (3) extracts the time-history of kinematic quantities as well as force/moment production due to any selected vortex on the immersed body. 

\begin{figure}
  \centerline{\includegraphics[scale=1.0]{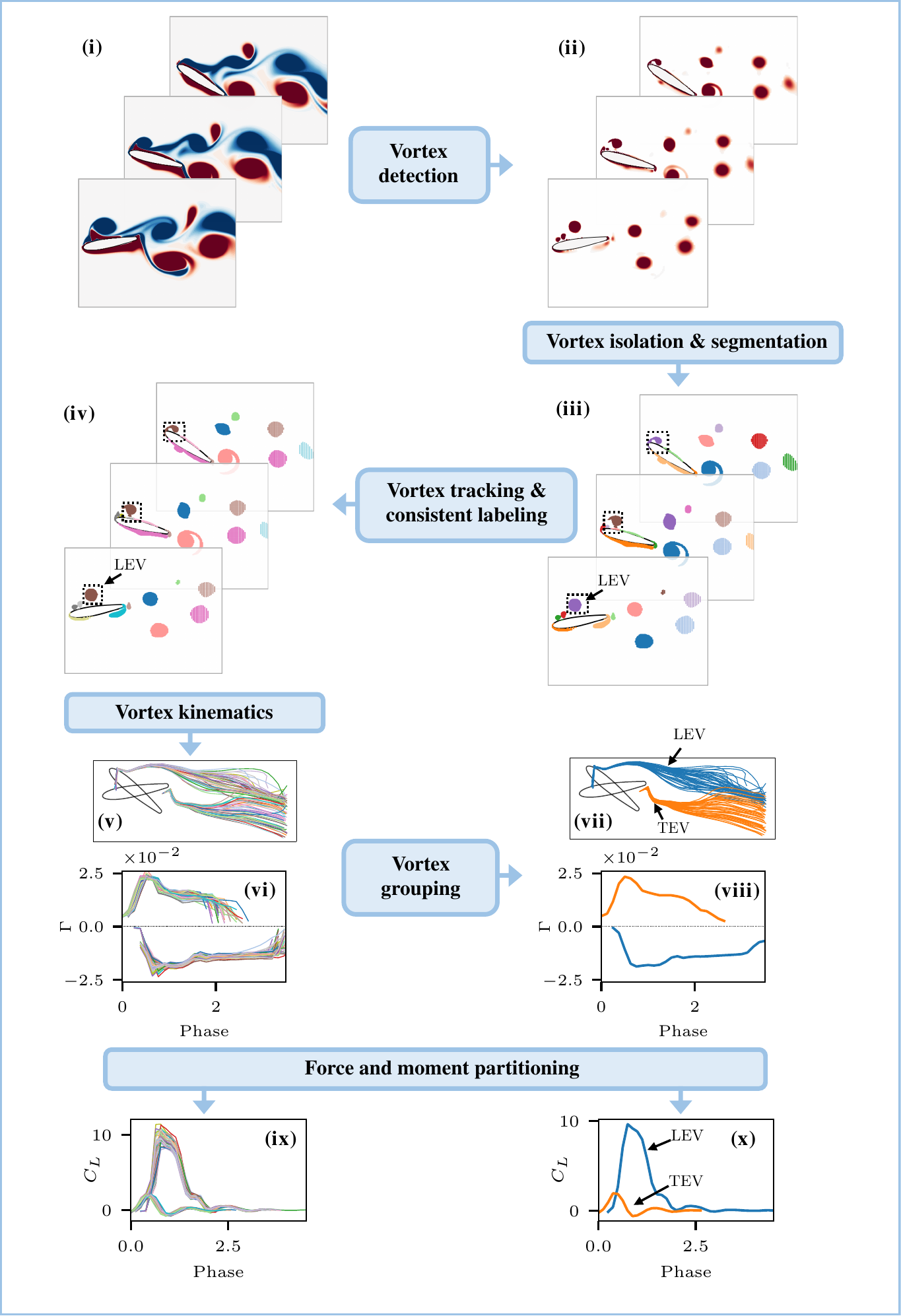}}
  \caption{Schematic of the steps involved in the framework for the automated tracking of vortices and estimation of aerodynamic loads. The steps are illustrated using a sample case of a sinusoidally pitching airfoil with dimensionless frequency $f^*=0.55$ and amplitude $A_\theta=25^{\circ}$.}
\label{fig:tracking_schematic}
\end{figure}

The input to this framework is time-resolved data of the fluid velocity field, as well as information on the position and velocity of all material points on the surfaces of the immersed body and external domain boundaries. Here we assume that the velocity field is defined within the fluid domain $V_f$, and is represented in discrete form, i.e. on a spatial grid using the set of grid points $\{\zeta\}$ and discrete temporal snapshots at $N_t$ time-instances. In this discussion, we will denote individual time-snapshots as $t_j$ where the temporal index is $j \in \{1,2,\cdots,N_t\}$. We will outline the steps involved in this framework using a sample case of a two-dimensional, sinusoidally pitching airfoil, with dimensionless oscillation frequency $f^* = fC/U_{\infty}=0.55$ and amplitude $A_{\theta}=25^{\circ}$. A schematic of the steps involved in this methodology is shown in figure \ref{fig:tracking_schematic}, where we show snapshots of results from each stage in the procedure for this sample case. 

\subsection{Vortex detection}
\label{sec:vortex_detection}

The first step in isolating and quantifying the influence of individual vortex structures is the task of detecting coherent vortex structures in the given flow-field. This problem of vortex detection has attracted significant research interest, and numerous definitions of coherent vortex structures have been proposed in literature, based on Eulerian \citep{Hunt1988EddiesFlows,Chong1990AFields,Jeong1995OnVortex} and Lagrangian \citep{Haller2015LagrangianStructures,Hadjighasem2017ADetection} criteria. In the context of this work, the aim of this vortex detection step is simply to identify the spatial volume corresponding to vortical regions in the flow-field, and this step can be implemented with many of the available vortex detection methods. In principle, the only requirement on the choice of the vortex detection method is that it generate an Eulerian scalar field representing the \textit{interior} of the region occupied by vortical structures. This requirement stems from the fact that estimating the loading induced by vortex structures, using the vorticity-induced force and moment shown in equation \ref{eq:vif_combined_approx}, requires the identification of the spatial volume corresponding to these structures. In fact, the majority of existing vortex detection criteria do satisfy this requirement, and can therefore be used as a starting point in this framework.

In the implementation presented here, we utilize the $Q$-criterion \citep{Hunt1988EddiesFlows} for detecting vortical regions in a given flow-field, primarily due to the formulation of the FMPM where $Q$ makes an explicit appearance. This results in the $Q$ scalar field, which is defined as in equation \ref{eq:q-definition}. This $Q$ field is a metric that compares the strength of rotation with strain in a local region of the flow, with  `vortices' corresponding to regions where $Q>0$ i.e. where rotation dominates over strain. In figure \ref{fig:tracking_schematic}, we show a sample vortex detection step for a 2D pitching airfoil. Figure \ref{fig:tracking_schematic}(i) shows the vorticity field at three time snapshots during the pitch-down motion for this case. These flow-fields represent the input data for this procedure, and figure \ref{fig:tracking_schematic}(ii) shows the $Q > 0$ field, which represents the vortices in the flow-field. 

\subsection{Vortex isolation and segmentation}
\label{sec:vortex_isolation}
Having identified vortical regions in the flow-field, the next step in this framework involves isolating the spatial volume corresponding to vortices, and then segmenting out the region occupied by each vortex. As shown in figure \ref{fig:tracking_schematic}(ii), the vortex detection step results in a scalar field representing the distribution of the chosen quantity (such as $Q$ in this case) at every time-snapshot. 
%
%
Using this scalar field, we perform a two-step method to isolate and segment the volume occupied by individual vortex structures. The first sub-step, i.e. the vortex isolation sub-step, simply involves applying an appropriate threshold to extract volumes within the domain that are occupied by vortices at time-snapshot $t_j$. In the context of the $Q$-criterion, we apply a threshold $Q > \delta$, where $\delta$ is small in order to approximate the more physically relevant $Q>0$ threshold for vortex detection. In general, the chosen threshold depends on factors such as errors/uncertainties in the underlying data, and on the vortex detection method used. In this work we use the threshold $\delta = 5$. We show in \ref{app:supplementary_results} that the aerodynamic loading due to flow structures isolated using $\delta = 5$ is very similar to that using $\delta = 0$. In terms of the flow-field data represented on the set of grid points $\{\zeta\}$, this thresholding results in a subset of grid points, $\{\zeta^\Omega\}= \{ \zeta : Q({\zeta}) > \delta \}$, which consists of several disconnected (i.e. isolated) vortical regions.

Next, we segment these vortical regions into distinct volumes occupied by each individual vortex. This is performed by utilizing the well-known DBSCAN (density-based spatial clustering of applications with noise) clustering algorithm \citep{Ester1996ANoise} on the set of grid points $\{\zeta^\Omega\}$. We note that while the use of clustering in the detection of coherent structures has been proposed in prior studies \cite{Froyland2015AData,Ser-Giacomi2015FlowTransport,Hadjighasem2016Spectral-clusteringDetection,Schlueter-Kuck2017CoherentTheory}, the aim here is not to \emph{detect} vortex structures using this approach. Rather, the use of clustering here simply aims to segment multiple distinct vortical regions in a flow and can be used in conjunction with other methods of vortex detection.

DBSCAN is a density-based clustering technique that detects clusters as groups of points that have at least a certain number of neighbouring points, $n_d$, within a specified distance, $\epsilon$. The required parameters, $\epsilon$ and $n_d$, are the maximum distance between two data-points for them to be considered in the same cluster, and the minimum number of data-points in the neighbourhood of another point to create a cluster, respectively. We find that $\epsilon$ can be conveniently chosen based on the maximum grid-spacing in the region of interest. Correspondingly, $n_d$ depends on the number of grid-points that surrounds any given point within $\epsilon$ distance. The result of the DBSCAN clustering at any time-snapshot $t_j$ is a segmentation of the domain into $N_j$ non-intersecting clusters, where each cluster corresponds to a distinct subset of the $\{\zeta^\Omega \}$ grid points. The $p^{\text{th}}$ cluster at time-snapshot $t_j$ is denoted by $V^j_p \subset \{\zeta^\Omega \}$, such that $V^j_p \cap V^j_q = \emptyset$, for $p \neq q$. Here the subscripts, $(p,q) \in \{1,\cdots,N_j\}$, are arbitrary numeric ``labels'' assigned to each cluster, and the superscripts specify the index for time-snapshot $t_j$. The grid points within each set $V^j_p$ therefore define the spatial extent of the $p^{\text{th}}$ vortical structure at any given time instant $t_j$. We denote the set of all such volumes at time-snapshot $t_j$ as $V^j_\Omega = \{   V^j_1,V^j_2,\cdots,V^j_{N_j} \} $.

A sample result of this procedure is shown in figure \ref{fig:tracking_schematic}(iii), where at each time-snapshot we see several clusters corresponding to distinct vortex structures. Each cluster is assigned a numeric label from 1 to $N_j$ at every time-snapshot (where $N_j$ itself can vary from one time-snapshot to the next). In figure \ref{fig:tracking_schematic}(iii), this label is graphically represented by the color of the vortex structure. This DBSCAN-based approach has several advantages in the context of clustering spatial regions in a flow-field. One is that the DBSCAN algorithm, unlike some other clustering techniques, does not require prior information on the number of clusters present in the flow domain. Secondly, this method is able to detect clusters of arbitrary shapes, which is particularly important in highly dynamic vortex-dominated flows. Lastly, a density-based technique that detects clusters based on the proximity of ``vortical'' grid points ($\zeta^\Omega$) to each other naturally lends itself to flow-field data. If the flow is sufficiently well-resolved to capture distinct vortical structures, it also necessarily has enough grid points between these vortical structures in order to distinguish them. This ensures that by isolating regions of $Q>\delta$, the vortex structures of interest will present themselves as disconnected ``dense'' spatial regions in the domain. These can be efficiently segmented using a density-based clustering technique.

\subsection{Vortex tracking and consistent labeling}
\label{sec:vortex_tracking_subsection}
Application of the above clustering algorithm independently at every time-snapshot often results in cluster labels (which, for time-step $t_j$ range from  $1$ to $N_j$) that have no temporal continuity with the other time snapshots. This is shown in figure \ref{fig:tracking_schematic}(iii), where the result of clustering is shown at three consecutive time-snapshots and numeric cluster labels are represented by different colors. We see, for instance, that the leading-edge vortex, which is highlighted using a dashed box at each time-snapshot in figure \ref{fig:tracking_schematic}(iii), is labelled (colored) differently between subsequent time-snapshots. It is easy to see that this mis-labelling also occurs for the other vortex structures that were isolated in the previous time-step. Consistent labeling of the vortex structures over all the time-snapshots is essential in order to generate continuous time-histories of any attributes associated with these vortex structures. These attributes may include the geometry of vortices (location, size, shape etc.) or kinematic/dynamical behaviour such as net circulation and crucially, the aerodynamic loads associated with the vortex, which are obtained from the FMPM.

To achieve temporally consistent labeling of the segmented vortices, we employ a vortex tracking procedure based on a simple model of vortex convection. We start by computing the centroids of all clusters identified at time $t_j$. This set of centroids is denoted by $\{   \vec{X}^j_1 , \vec{X}^j_2, \cdots, \vec{X}^j_{N_j}\}$. The centroids are obtained as an integrated average of the coordinates within each corresponding volume $V^j_p \in V^j_\Omega$. The numerical integration can done using various schemes and here we employ the simplest, midpoint (area-weighted) integration scheme. Similarly, we define the convection velocity of each cluster as an area-weighted average of the flow velocity within each set of grid points $V^j_p$. These are denoted by  $\{\vec{U}^j_1, \vec{U}^j_2 , \cdots , \vec{U}^j_{N_j}\}$. Then at any time $t_j$, we first compute the predicted position of all structures detected at the previous time-snapshot using a Forward Euler scheme. For the $p^{\text{th}}$ cluster in the previous time-snapshot, this predicted position is given by $\vec{X}^{\prime}_{p} = \vec{X}^{j-1}_p +  \vec{U}^{j-1}_p \Delta t$, where $p \in \{1,...,N_{j-1}\}$ denotes the cluster labels at time $t_{j-1}$ and $\Delta t$ is the time-step between these consecutive snapshots. Hence at every time-snapshot $t_j$, we have the actual centroidal positions of each vortex structure detected directly from the flow-field at that time-snapshot, i.e. $\{\vec{X}^j_1, \cdots, \vec{X}^j_{N_j}\}$, as well as predicted positions of all vortex structures computed from the previous time-snapshot, i.e. $\{   \vec{X}^\prime _1,\cdots,\vec{X}^\prime_{N_{j-1}}\}$. 

Subsequently, a distance matrix of size $(N_j \times N_{j-1})$ is computed between the actual vortex centroids at $t_j$ and the centroids of the vortex structures predicted from the previous time-snapshot. This matrix is denoted as $D$, and its $p^{\text{th}}$-row and $q^{\text{th}}$-column entry is given by $D_{p,q} = | \vec{X}^j_p - \vec{X}^\prime_q | $. Here $|\cdot|$ is the $L$-2 norm. In order to improve the robustness of the distance comparison, we also include the sign of the average vorticity in each cluster as an extra dimension in this difference. These pairwise distances are then sorted in ascending order of magnitude, the rationale being that small entries in $D$ likely correspond to the same physical vortex structure between successive snapshots. Each pair of labels, $(p,q)$ in this sorted list is tested against certain conditions, such as either the label in the present or previous time step has not been already assigned to another cluster, or that $D_{p,q}$ is not greater than a specified distance. For pairs of labels that satisfy these conditions the label at the current time-snapshot ($p$) is changed to match that at the previous time-snapshot ($q$). This is repeated until either all cluster labels in the current time-snapshot have been matched with a corresponding cluster from the previous time-snapshot, or there are no more pairs that satisfy the specified criteria. The remaining cluster labels at the current time-snapshot are then considered as being vortices that have appeared at the current time-step, and are given new labels that are distinct from all previously used labels. The labels from the previous snapshot ($t_{j-1}$) that do not find a match with any vortices at $t_j$ correspond to structures that have either exited the domain of interest or dissipated below the prescribed threshold. The labels corresponding to these structures are retired. The final result is that the label for each identified vortex is carried from one time to the next in a continuous and consistent manner. Furthermore, the time-varying volume occupied by a vortex with label $p$ is now fully described at any time $i$ by the set of grid points within each $V^j_p$.

In figure \ref{fig:tracking_schematic}(iv) the result of this vortex tracking process is shown. We see that the leading-edge vortex, highlighted using the dashed box at each time-snapshot, as well as other vortices which were arbitrarily labelled at each time-snapshot in figure \ref{fig:tracking_schematic}(iii) are now labeled in a way that results in the consistent tracking of these vortex structures. We note that the above procedure for feature tracking falls under the category of ``attribute correspondence'' in the flow visualization domain (where the attribute in this case is the position), and a similar method has been shown to work well even in 3D flows \citep{Reinders2001VisualizationDetection,Post2003TheTracking}.

\subsection{Vortex kinematics, ranking and grouping}
\label{sec:categorizing_vortex}
Once each segmented vortex has a unique and consistent label over time, we can compute temporal histories of various kinematic attributes for these vortices including, but not limited to, 
\begin{itemize}
\setlength\itemsep{0em}
    \item Centroid location, including location of inception and location of exit.
    \item Vortex area (2D) or volume (3D).
    \item Shape (as defined by the vortex boundary).
    \item Vortex strength.  
\end{itemize}
Additionally, vortices can now also be ranked based on any of these kinematic attributes. Here we choose to rank the vortices in terms of their circulation (representing vortex strength). For a vortex structure with label $p$, we compute its time-dependent circulation (in 2D) as $\Gamma_p(t_j) = \sum_{V^j_p} \omega_z \Delta x \Delta y$, where the summation is over the grid cells in the set $V^j_p$ at each time-snapshot ($t_j$), $\omega_z$ is the time-varying vorticity at the center of each grid cell, and $\Delta x$ and $\Delta y$ are the linear dimensions of the grid cell. Due to the periodic nature of the pitching airfoil problem being analyzed here, we perform this ranking over each cycle of the airfoil's oscillation. This is done by integrating the magnitude of circulation for each vortex detected in a particular cycle over all time-snapshots within that cycle. We then sort the vortices in every cycle by this total magnitude of circulation, and retain only the top few vortices for analysis. It must be noted that this rank-reduction of the set of isolated vortices can be performed in several different ways, and is only intended to simplify subsequent analysis.


In figures \ref{fig:tracking_schematic}(v) and \ref{fig:tracking_schematic}(vi) we show sample results of the kinematic analysis described above. Figure \ref{fig:tracking_schematic}(v) shows vortex trajectories, represented as the loci of the centroid of each structure, for the two ``top-ranked'' vortices identified in each oscillation cycle using the circulation-based ranking described above. Figure \ref{fig:tracking_schematic}(vi) shows time-series plots of circulation for these two ``top'' vortices, plotted against the phase of the airfoil's oscillation. It is clear that the trajectories of the two strongest vortices, although they seem chaotic, all correspond to vortices shedding off either the leading or trailing edge of the airfoil. In fact, the leading-edge vortex (LEV) and trailing-edge vortex (TEV) are observed to be the strongest vortices, in terms of circulation, for the majority of cases analyzed in the current data set. This is evident from the flow snapshots in figures \ref{fig:tracking_schematic}(i)-(iv) for the specific case being analyzed in this section.

In addition to the analyses suggested above, the available kinematic information can also be used to group the set of the detected vortex structures based on similarities in any chosen attribute. 
For instance, the centroid location of each vortex as a function of time allows us to group vortices based on location of inception. In particular, in the rank-reduced set of vortices obtained above, vortices emerging from the leading-edge could be assigned to one group and those emanating from the trailing-edge could be assigned to a different group. 
We perform this grouping of vortex structures by using a clustering-based approach. We again use the DBSCAN algorithm described above for this task, primarily because it does not require knowledge of the number of classes beforehand. Figures \ref{fig:tracking_schematic}(vii) and \ref{fig:tracking_schematic}(viii) show the result of such a vortex grouping, applied to the two top-ranked vortices identified using circulation. The identified groups are shown using different colors for each group, and figure \ref{fig:tracking_schematic}(vii) shows the categorized spatial trajectories. Here the grouping is based on the location of the vortex centroid at its inception, its location on exiting the domain of interest, and the mean and standard deviation of its circulation. We see that this method works well even when the kinematics are not perfectly repeatable over each cycle. Further, ensemble-averages can computed over a given group, and figure \ref{fig:tracking_schematic}(viii) shows the ensemble-averaged circulation as a function of oscillation phase for two groups of vortices, the LEVs and the TEVs.

\subsection{Force and moment partitioning}
\label{sec:dynamical_analysis}
Having accomplished the isolation and tracking of the various vortex structures, we can now perform the final step in this analysis framework, which is to quantify the aerodynamic loading due to these flow structures on the immersed body. As shown in equation \ref{eq:vif_combined_approx} of section \ref{sec:fpm}, the induced force and moment on an immersed body due to vortical regions of the flow can be quantified by integrating kinematic flow-field quantities over specifically constructed integral volumes. Thus, for a vortex structure with label $p$, which occupies the volume specified by $V^j_p$ at time $t_j$, the time-varying induced loads can be estimated by the following expression:
\begin{equation}
    {C^{\omega_p}_{F_i}} (t_j) \approx  - 2 \int_{V^j_p} Q \; \phi_{i} \; dV \quad ; \quad 
    {C^{\omega_p}_{M_k}} (t_j)  \approx  - 2 \int_{V^j_p} Q \; \psi_{k} \; dV 
    \label{eq:vif_vortex}
\end{equation}
Here the integrands  can be readily calculated at every time-snapshot given the instantaneous flow-field and geometry of the immersed surface $B$. Therefore, using the procedure described above to isolate and track the volumes corresponding to each vortex structure in the flow makes it straightforward to compute the force and moment induced by each of these structures as they evolve with the flow 

In figures \ref{fig:tracking_schematic}(ix) and \ref{fig:tracking_schematic}(x) we show a result of this force computation for the sample case discussed here, where the coefficient of lift ($C_L$) due to the two ``top-ranked'' vortices, i.e. the LEV and TEV, are plotted against the phase of the oscillation. The multiple curves in figure \ref{fig:tracking_schematic}(ix) correspond to LEVs and TEVs shed during several oscillation cycles of the pitching airfoil. This procedure therefore allows us to quantify the force and moment production due to each of these vortices. Further, the vortex grouping procedure described above also allows the computation of statistics of the aerodynamic loads over multiple occurrences of a single type of vortex. This is demonstrated in figure \ref{fig:tracking_schematic}(x), where the ensemble-averaged $C_L$, computed over all occurrences of LEVs and TEVs separately, is shown as a function of the phase of oscillation.
Thus, using the framework described in this section, we can analyze high-dimensional, time-resolved flow-fields and extract a wide range of quantities associated with the kinematics as well as dynamical influence of individual vortex structures.

\section{Application to pitching airfoils}
\label{sec:pitching_airfoil}

We now present an application of the methods described in the previous sections to the configuration which was introduced earlier in figure \ref{fig:vortex_intro}: the two-dimensional flow past an  airfoil which is undergoing prescribed sinusoidal pitch oscillations over a range of amplitudes and frequencies of oscillation. As we will show, this canonical problem exhibits numerous distinct vortex-dynamic regimes, which have been shown to be more complex \citep{Schnipper2009VortexFoil} than those behind the more well-studied problem of an oscillating cylinder. Further, as highlighted in figure \ref{fig:vortex_intro}, the flow is characterized by several dominant interacting vortex structures, and even relatively small changes in the kinematics lead to substantially different vortex-dynamic behaviour. In this section, we will first employ a data-driven method to reduce this large ensemble of flow-fields to a small number of distinct regimes, and subsequently utilize the methods discussed in prior sections to analyze these vortex-dominated flow-fields at the level of individual vortices.


The data-set analyzed here consists of 165 distinct cases of a two-dimensional NACA0015 airfoil, which undergoes prescribed sinusoidal pitching oscillations at Reynolds number, defined based on free-stream velocity $U_{\infty}$, and chord-length $C$, as $Re = U_{\infty} C/ \nu = 1000$. The sinusoidal pitching amplitude is given by $\theta = \theta_0 + A_{\theta}\sin{(2 \pi f^* t)}$. Here $\theta$ is the instantaneous angle-of-attack, $\theta_0$ is the mean angle-of-attack, and $A_{\theta}$ is the pitching amplitude (reported in degrees). Further, $f^*$ is the dimensionless pitching frequency, given by $f^* = f C/U_{\infty}$, with corresponding oscillation period $T^*=1/f^*$, and $t$ is the dimensionless time (also non-dimensionalized by $U_{\infty}$ and $C$). The airfoil is forced to pitch about mid-chord with mean angle-of-attack $\theta_0  =15^{\circ}$. We analyze oscillation amplitudes and frequencies in the range $0.5 \leq A_{\theta} \leq 40$ and $0.10 \leq f^* \leq 0.70$ respectively. The time-resolved flow-field data used for this analysis consists of $500$ temporal snapshots for each case, recorded after the flow has reached a stationary state. This corresponds to between $10$ and $70$ oscillation cycles, depending on $f^*$ for the particular case. The flow-field data is generated via computational simulations, using a sharp-interface immersed boundary method \citep{Mittal2008ABoundaries,Seo2011AOscillations}. For more details of the simulations performed, the reader is referred to ref. \citep{Menon2019}. The force and moment partitioning of section \ref{sec:fpm} is implemented in the framework of this sharp-interface immersed boundary method, using second-order accurate discretization in space and time. The Laplace equations corresponding to the auxiliary potential fields, equations \ref{eq:fpm_scalar}, are solved using a geometric multigrid method.

\subsection{Rank-reduction based on vortex shedding regimes}
\label{sec:airfoil_wake}
\begin{figure}
  \centerline{\includegraphics[scale=1.0]{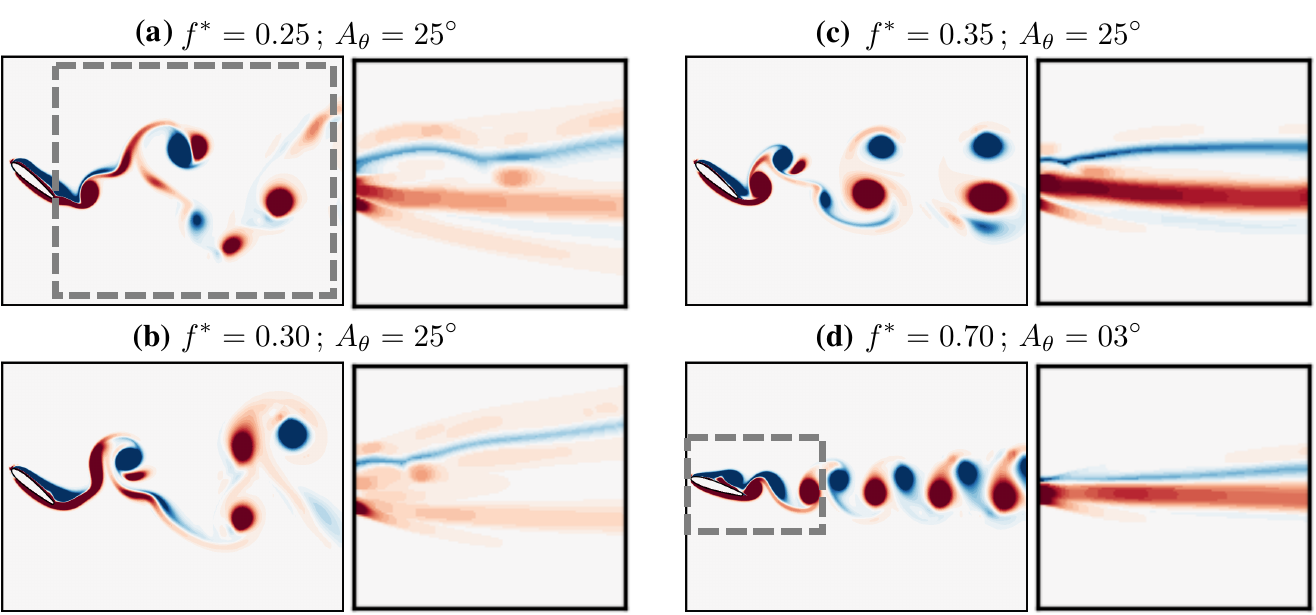}}
  \caption{Instantaneous snapshots of the vorticity field (left panel for each case) for four representative cases of pitching airfoils, and corresponding time-averaged vorticity fields in the wake ($\bar{\omega}$; right panel for each case). The instantaneous snapshots are shown at the phase of maximum angle of attack, in order to highlight leading and trailing edge vortex interactions and trajectories. (a) $f^*=0.25$, $A_\theta=25^{\circ}$; (b) $f^*=0.30$, $A_\theta=25^{\circ}$; (c) $f^*=0.35$, $A_\theta=25^{\circ}$; (d) $f^*=0.70$, $A_\theta=03^{\circ}$. The dashed box in (a) shows the region used for calculating $\bar{\omega}$, and that in (d) shows the domain used for the analysis in section \ref{sec:airfoil_vortex_tracking}.}
\label{fig:snapshots}
\end{figure}
\begin{figure}
  \centerline{\includegraphics[scale=1.0]{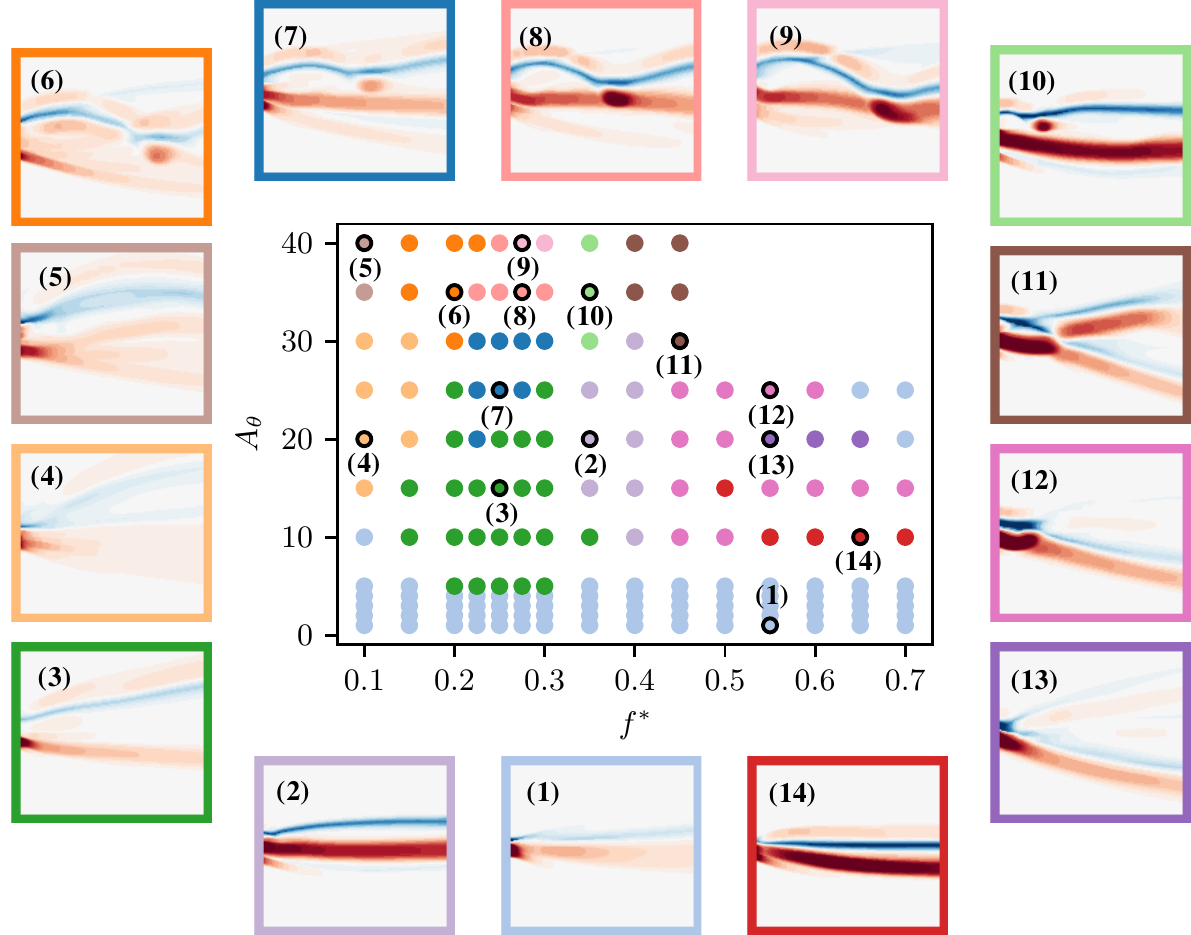}}
  \caption{Final clustering result for distinct vortex-dynamic regimes in the data-set of pitching airfoil simulations. The plot in the center shows all the simulations in the data-set in $f^*$-$A_{\theta}$ space, with markers of different colors corresponding to different clusters of vortex-dynamic regimes. Time-averaged vorticity fields corresponding to the simulation closest to the centroid of each cluster (indicated using numeric labels for each cluster) are shown around the central $f^*$-$A_{\theta}$ plot. }
\label{fig:clust_result}
\end{figure}

Due to the large size of the present data-set, we begin by performing a ``rank-reduction'' of this data to facilitate the analysis. The aim here is to reduce the set of 165 time-resolved flow-fields to a smaller set of distinct regimes (i.e. a ``rank-reduced set'') that captures the important features present in the data set. We then perform the subsequent vortex analysis on this rank-reduced set of cases. In the present application, distinct vortex-dynamic regimes are identified based on vortex patterns in the wake of the pitching airfoil. Further, we take a data-driven approach to perform this rank-reduction, by identifying groups (or clusters) within this ensemble of flow-fields that have similar vortex-wake patterns. 

Given this large ensemble of time-resolved flow-fields representing each member of the ensemble, the first step involves extracting appropriate information from each case in order to identify similarities in their vortex patterns. While there are several ways to extract important patterns from time-resolved flow-fields, such as modal decomposition techniques \cite{Taira2017ModalOverview}, we use a ``zeroth-order mode'' of the flow-field -- the time-averaged vorticity field. In figure \ref{fig:snapshots}, we show instantaneous snapshots of vorticity (left panel for each case) for four representative cases from this data-set, along with their corresponding time-averaged wake-vorticity fields (right panel for each case). It is interesting to note that, although the cases in figures \ref{fig:snapshots}(a)-(c) have similar kinematics, they show substantial differences in their vortex dynamics, and consequently in their mean vorticity patterns. The wake region used for this calculation of mean vorticity is shown using a dashed rectangular box in figure \ref{fig:snapshots}(a), and is of size $4.5C \times 4C$. Using this extracted ``feature vector'' of the flow-field for each case, we subsequently use a sequence of data-driven tools to discover similarities within the data-set, and eventually arrive at a small number of cases representing distinct vortex dynamic regimes. Further details of the method, which involves dimensionality reduction using principle component analysis, followed by clustering based on Gaussian mixture modeling, and a statistical evaluation of the robustness of the results, are provided in \ref{app:supplementary_results}.

The rank-reduction process employed here results in 14 clusters representing distinct vortex-wake patterns. Figure \ref{fig:clust_result} shows the demarcation of all cases in the data-set into these 14 clusters. The plot at the center of figure \ref{fig:clust_result} shows the frequency and amplitude of all the cases in the data-set using circles, and these circles are colored based on the cluster membership for each case. The case closest to the centroid of each of these 14 clusters is highlighted using a numeric cluster label in this frequency-amplitude. Further, we also show the mean vorticity field ($\bar{\omega}$) corresponding to the ``centroidal'' case for each cluster. These $\bar{\omega}$ plots are labelled using the same cluster labels as well as framed using the cluster color corresponding to each cluster in the frequency-amplitude plot. We see that this procedure results in the identification of several distinct vortex-dynamic regimes. For the remainder of this paper, We use these representative centroidal cases to analyze the vortex dynamics in this data-set.

\subsection{Analysis of vortex kinematics}
\label{sec:airfoil_vortex_tracking}

We now demonstrate the utility of the analysis framework described in section \ref{sec:vortex_tracking}, applied to the distinct vortex-dynamic regimes identified in the previous sub-section. For each representative case in figure \ref{fig:clust_result}, the time-resolved flow-field data is analyzed using the framework outlined in section \ref{sec:vortex_tracking} and in the discussion above. While the wake-identification process in the previous sub-section takes a ``global'' view of the flow-field, the aim of the methods demonstrated here is to dissect these regimes in terms of individual vortex structures -- their kinematics, dynamics and the forces/moments they induce on the pitching airfoil. 

\begin{figure}
  \centerline{\includegraphics[scale=1.0]{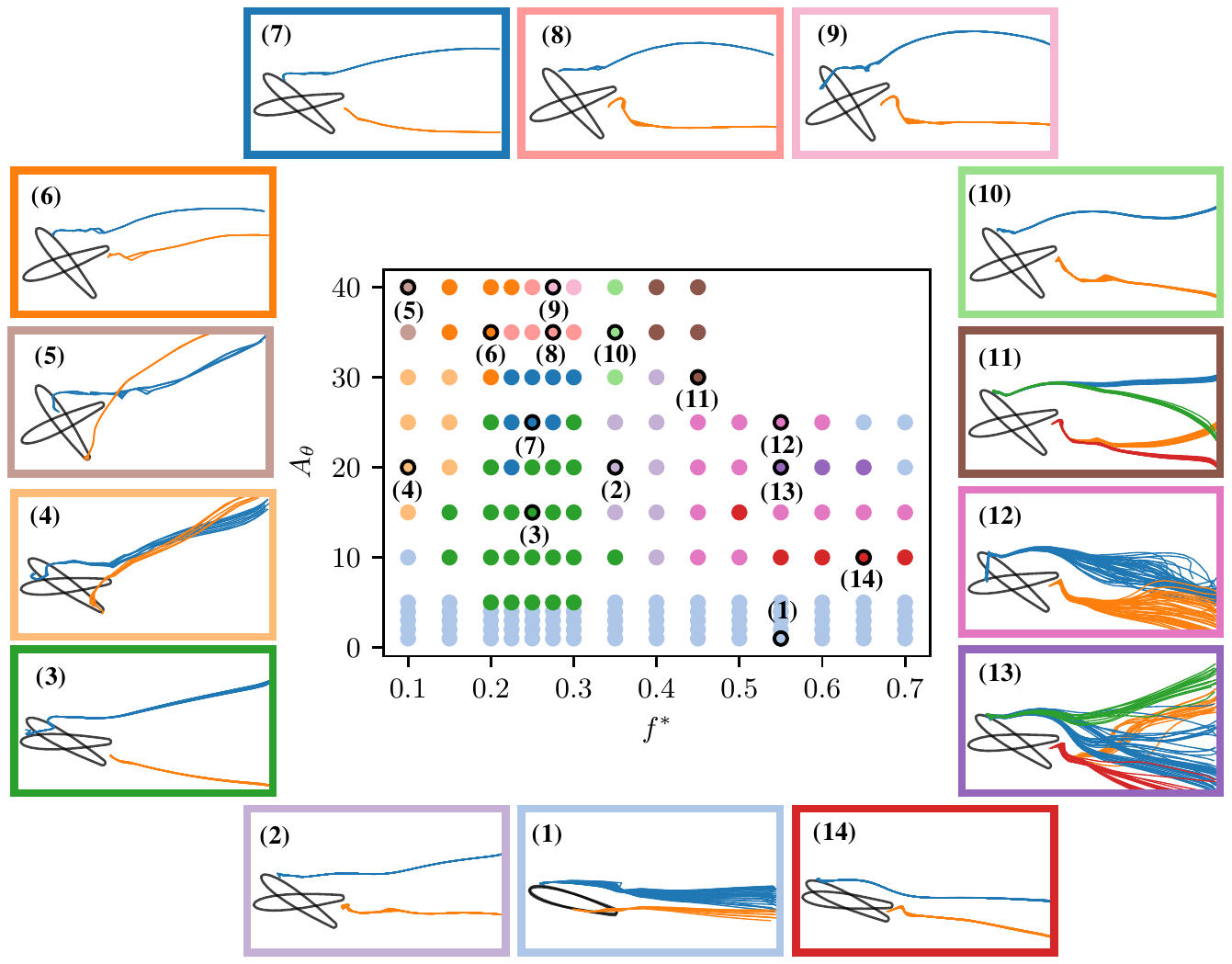}}
  \caption{Leading-edge vortex (\protect\thinblueline) and trailing-edge vortex (\protect\thinorangeline) trajectories corresponding to each vortex-wake regime identified in section \ref{sec:airfoil_wake}. Figure in the center shows all the simulations in the data-set in $f^*$-$A_{\theta}$ space, with markers of different colors corresponding to different regimes. The vortex trajectories shown correspond to the simulation closest to the centroid of each cluster. Also shown on each trajectory plot is the maximum and minimum angle-of-attack of the oscillating airfoil for each case.}
\label{fig:clust_traj}
\end{figure}
We will largely focus on the leading-edge vortex (LEV) and the trailing-edge vortex (TEV) in this discussion due the dominant role they play in the fluid dynamics of pitching airfoils. In addition, the LEV and TEV are seen to be amongst the ``strongest'' vortices (in terms of circulation) for all cases analyzed here. However, we note that the methods demonstrated can be applied to the simultaneous analysis of many other flow structures as well, and we present an example of such an analysis later in this section. Since a particular focus of this method is to evaluate the influence of specific vortex structures in terms of forces and moments induced on a body, we will center this analysis on the flow in the vicinity of the airfoil. The spatial domain of interest is hence constructed such that its horizontal extent begins upstream of the airfoil at the leading edge (when the airfoil is at $\theta = 0^{\circ}$) and extends $1.5C$ downstream of the trailing edge. The vertical size of the domain of interest is $2C$. A schematic of this analysis domain is shown in figure \ref{fig:snapshots}(d).

In figure \ref{fig:clust_traj} we show the trajectories of all the LEVs (in blue) and TEVs (in orange) for each of the 14 representative (``centroidal'') cases identified using the rank-reduction procedure described above. These trajectories consists of an average of $40$ oscillation cycles after the flow reaches a stationary state. As in figure \ref{fig:clust_result}, all simulations in the data-set are represented in the frequency-amplitude plot using circles corresponding to their frequencies and amplitudes of oscillation, colored by their cluster membership. The plots showing LEV and TEV trajectories for each of the 14 representative (``centroidal'') cases are given numeric labels corresponding to each cluster's  ``centroidal'' case indicated in the frequency-amplitude plot. Further, these LEV and TEV trajectories are overlaid on snapshots of the airfoil surface at its extreme angles-of-attack for each case. 

We see from figure \ref{fig:clust_traj} that the distinct wake regimes identified previously also correspond to several distinct LEV and TEV trajectories in the vicinity of the airfoil. These vortex trajectories plots reveal several interesting features of the vortex dynamics in this problem. First, for all but two cases the LEV is shed at the maximum pitch-up phase and the TEV is shed at maximum pitch-down phase. The two cases labels $(4)$ and $(5)$ for which this does not happen are both the lowest frequency cases in the current data set ($f^*=0.10$). These two cases also show trajectories that are quite distinct; whereas in most other cases (excluding cases $11-13$) the LEV and TEVs travel along trajectories that do not cross, these two cases show trajectories that cross in the near wake and are indicative of vortex entrainment. As we will show, these observations are significant in terms of the forces induced by the TEV as the oscillation frequency is varied.  

A second interesting behavior that is noted is that for case $(11)$, where we see a bifurcation in the trajectories of the LEVs and TEVs. Examination of the vortex shedding behavior for this case shows that this is associated with the appearance of period-doubling bifurcation, wherein shed vortices alternate between these two trajectories from cycle to cycle. Furthermore, examination of cases $(10)-(13)$ also reveals what appears to be a period-doubling route to chaos in this flow. For case $(10)$, the trajectories are highly repeatable from cycle-to-cycle. A slight increase in frequency coupled with a decrease in amplitude results in the period-doubling behavior seen for case $(11)$. Further increase in frequency and decrease in amplitude, in case $(12)$, shows the appearance of cycle-to-cycle variability in the vortex trajectories which can be viewed as emerging chaotic behavior. Case $(13)$ shows further increase in non-periodicity of the flow with the vortex trajectories of both the LEVs and TEV showing large cycle-to-cycle variations, which is indicative of well-developed chaos. 

\begin{figure}
  \centerline{\includegraphics[scale=1.0]{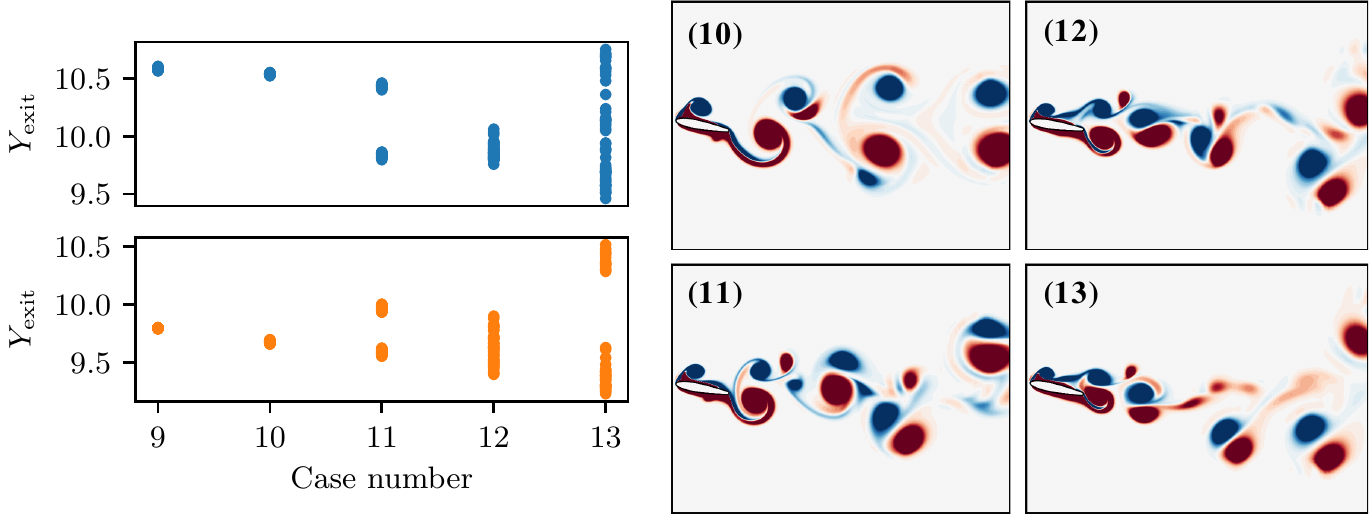}}
  \caption{Scatter plot on the left shows the $Y$-coordinate of each LEV (top panel; \protect\bluecircle) and TEV (bottom panel; \protect\orangecircle) as it exits the analysis domain, for cases $(9)$-$(13)$ in figure \ref{fig:clust_traj}. The scatter plot is generated from vortices in $\approx 40$ oscillation cycles in each case. Right panel shows an instantaneous snapshot of vorticity for the cases numbered $(10)$-$(13)$ in figure \ref{fig:clust_traj}.}
\label{fig:vort_bifurcation}
\end{figure}
This bifurcation in trajectories is highlighted in figure \ref{fig:vort_bifurcation}, where the scatter plot on the left shows the $Y$-coordinate of each LEV (blue circles; top panel) and TEV (orange circles; top panel) as it exits the analysis domain. These exit locations are plotted against the case numbers used in the above discussion. We see that cases $(9)$ and $(10)$ each show one repeatable exit location each for the LEV and TEV, whereas the period-doubling in case $(11)$ shows up as two repeatable exit locations. This exit location gets less repeatable for cases $(12)$ and $(13)$, with case $(13)$ in particular showing a large spread in exit locations. This resembles a typical return map for such period-doubling bifurcations, and a period-doubling route to chaos has indeed been observed in previous work on unsteady airfoils \citep{Sarkar2008NonlinearVibration}. However, the occurrence of such a bifurcation is not obvious from simple flow-field visualizations, particularly in very large data-sets such as this. This is highlighted in the right panel of figure \ref{fig:vort_bifurcation}, where we show instantaneous vorticity snapshots for the cases $(10)$-$(13)$. While these snapshots show some differences in the flow, they are not indicative of clear trends in the kinematics such as that discussed above. Therefore the ability to identify such behaviour in very large ensembles of flow data is a demonstration of the utility of the current data-driven analysis approach. 

\begin{figure}
  \centerline{\includegraphics[scale=1.0]{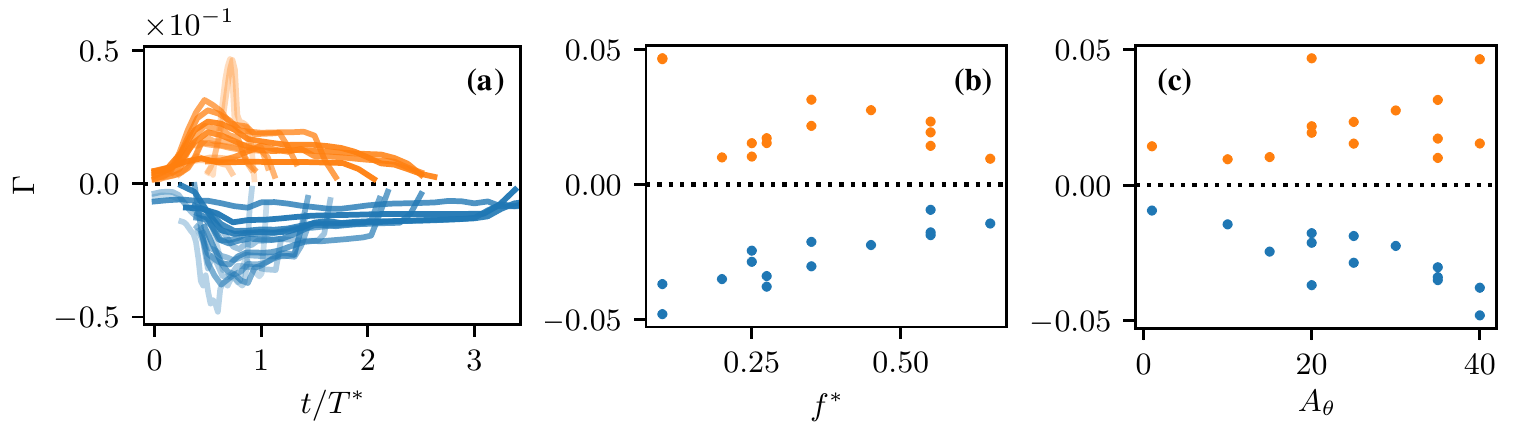}}
  \caption{(a) Phase-averaged circulation, $\Gamma$, for LEVs (\protect\blueline; $\Gamma < 0$) and TEVs (\protect\orangeline; $\Gamma > 0$), plotted against oscillation phase ($t/T^*$) for all 14 cases representing distinct vortex dynamics; (b) Scatter plot of peak $\Gamma$ versus oscillation frequency ($f^*$) for LEVs (\protect\bluecircle) and TEVs (\protect\orangecircle); (c) Scatter plot of peak $\Gamma$ versus oscillation amplitude ($A_{\theta}$) for LEVs (\protect\bluecircle) and TEVs (\protect\orangecircle). Peak value is defined as most negative value for LEV and most positive value for TEV. Note: Transparency in (a) is arbitrary and is purely for visualization.}
\label{fig:circulation}
\end{figure}

In addition to vortex trajectories, the method allows us to extract a variety of other quantities associated with the kinematics and geometry of the vortices, including, for instance, their circulation. As an example, figure \ref{fig:circulation}(a) shows a plot of phase-averaged circulation versus oscillation phase ($t/T^*$) for LEVs (\blueline; blue) and TEVs (\orangeline; orange) in all 14 cases representing distinct vortex dynamics identified above. The circulation is computed as $\Gamma = \int_{A_{\Omega}} \omega_z dA$, where $A_{\Omega}$ is the area occupied by a particular vortex and $\omega_z$ is the out-of-plane vorticity. We overlay this circulation time-series for all 14 cases to highlight the wide variety of observed behaviour. Further, this data can also be parsed to analyze trends in circulation versus airfoil kinematics. Figures \ref{fig:circulation}(b) and \ref{fig:circulation}(c) show the peak value of circulation (defined as maximum value for TEV and minimum value for LEV) plotted against $f^*$ and $A_{\theta}$ respectively. We see that clear trends emerge from this data, showing that the magnitude of peak LEV circulation is inversely proportional to $f^*$ and directly proportional to $A_{\theta}$. The TEV shows an interesting non-monotonic trend against $f^*$ and is weakly monotonic with $A_{\theta}$. While in the current paper we do not focus further on this data, we point out that this easily extracted time-resolved circulation data could be useful for understanding vortex dynamics as well as for the development and validation of simplified vortex models \citep{Wang2013Low-orderFormation,Ramesh2015Limit-cycleShedding}.

\begin{figure}
  \centerline{\includegraphics[scale=1.0]{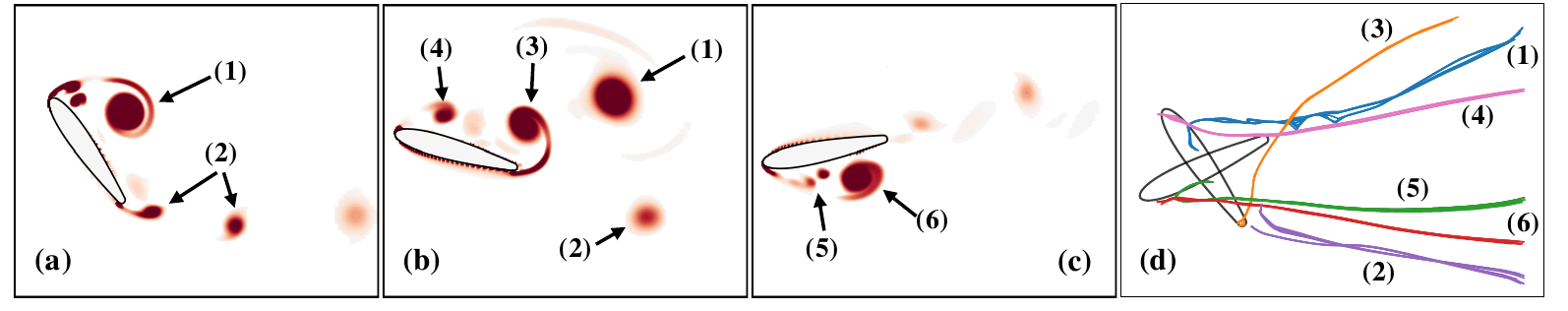}}
  \caption{An example showing the use of current framework to isolate, track, and group several ($>2$) vortices simultaneously. For this case, $f^*=0.10$ and $A_{\theta}=40^{\circ}$. (a)--(c) Instantaneous snapshots of the flow visualized using regions of $Q > 0$. Snapshots are shown at three phases in a single oscillation cycle. Arrows and numeric labels highlight six types of vortices that are analyzed. (d) Spatial trajectories of the six types of vortices highlighted in the flow snapshots. Colors show different vortex types and numeric labels correspond to those in (a)--(c).}
\label{fig:top_clust_0.10_40}
\end{figure}

As a final demonstration of this framework for kinematic analyses, we present an example of its application to many simultaneously evolving and interacting vortex structures, in addition to the LEV and TEV. As mentioned earlier, the discussion thus far focused exclusively on the LEV and TEV due to the fact that they are the dominant vortices in the majority of cases. However for oscillations at very low frequencies, we see the generation and shedding of several other vortices that are comparable to, or only slightly weaker than the LEV and TEV in terms of circulation. Hence we choose the case with $f^*=0.10$ and $A_{\theta}=40^{\circ}$ for this demonstration. Figures \ref{fig:top_clust_0.10_40}(a)-(c) show three instantaneous snapshots of the vortices in the flow-field, visualized as regions of $Q > 0$. Several vortices of varying strength and size have been indicated using numeric labels in these snapshots. We see that this case features a complex flow-field with vortices shed off the upper and lower surfaces of the leading edge, labeled $(1)$ and $(6)$ respectively. Further, there are also secondary LEVs shed off both these surfaces, labeled $(4)$ and $(5)$ respectively, as well as multiple types of trailing-edge vortices, labelled $(2)$ and $(3)$ respectively. Figure \ref{fig:top_clust_0.10_40}(d) shows the extracted vortex trajectories and and we see that the method is able to accurately detect, track, as well as group all these different vortices. This is also an example of a case where identifying groups of vortices based purely on location of inception would not be sufficient, and using other attributes that are available with the current method (such as mean and standard deviation of circulation in this case) becomes necessary. Although not shown here for brevity, the accurate isolation, tracking, and grouping of these vortices now allows us to extract and analyze a variety of kinematic and dynamic quantities relevant to these vortices, in much the same way as in the analysis of LEVs and TEVs presented in the rest of this paper.

\subsection{Vortex-induced forces and moments}

\begin{figure}
  \centerline{\includegraphics[scale=1.0]{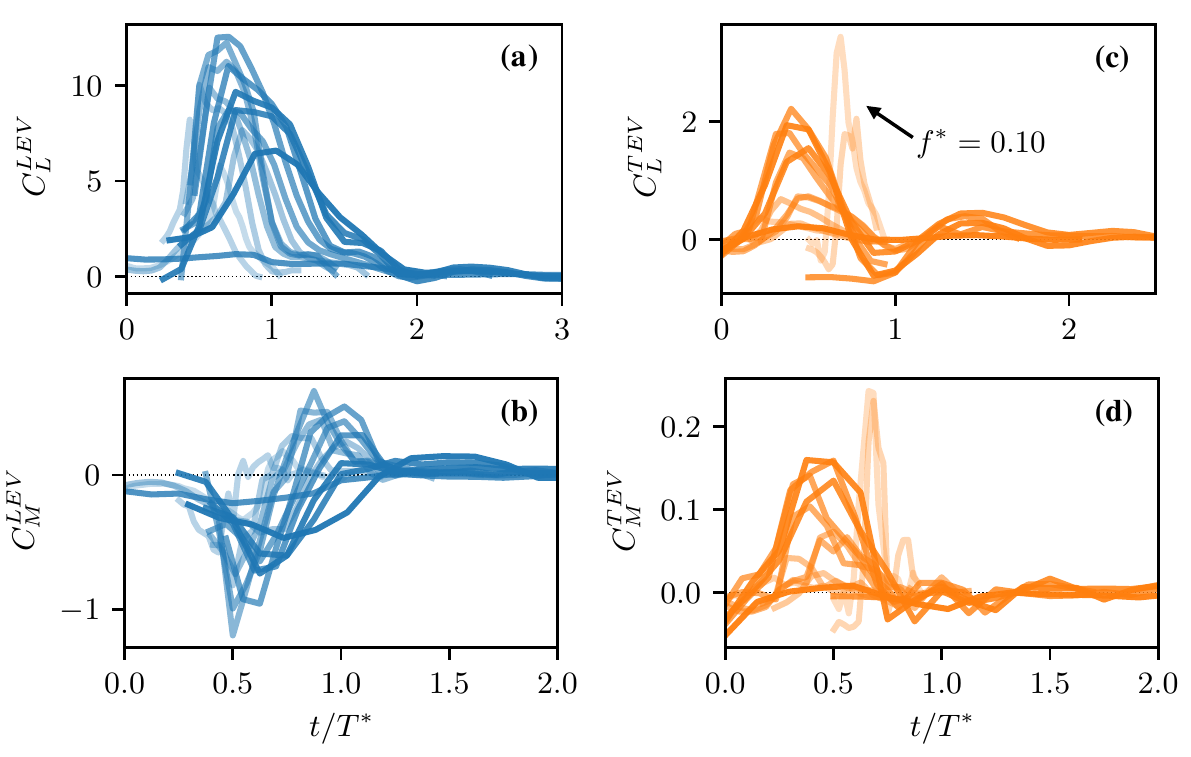}}
  \caption{Phase-averaged $C_L$ (top panel) and $C_M$ (bottom panel) induced by LEVs (\protect\blueline) and TEVs (\protect\orangeline), overlaid for all 14 cases representing distinct vortex dynamic regimes. $X$-axis shows the phase of oscillation ($t/T^*$); (a) $C_L$ due to LEVs; (b) $C_M$ due to LEVs; (c) $C_LM$ due to TEVs; (d) $C_M$ due to TEVs. Note: Transparency is arbitrary and is purely for visualization.}
\label{fig:cm_cl_all}
\end{figure}

We now demonstrate the utility of this framework in dissecting the dynamical influence of individual flow-structures on the immersed surface. As in the previous section, the focus here will be on the influence of the LEV and TEV. We begin with an overview of the force and moment generated by the LEV and TEV in all 14 centroidal cases. This is shown in figure \ref{fig:cm_cl_all}, where the phase-averaged $C_L$ and $C_M$, overlaid on each other for all 14 cases, are plotted against the phase of oscillation, $t/T^*$. Figures \ref{fig:cm_cl_all}(a) and \ref{fig:cm_cl_all}(b) show the LEV-induced coefficients of lift ($C_L^{LEV}$) and moment ($C_M^{LEV}$) respectively, and we see that there is a wide range in the peak value of lift and moment induced by the LEV as the airfoil's oscillation kinematics are varied. Further, we see clear variations in the phase of the peak lift and moment, which is an important consideration and will be discussed in more detail in this section. Figures \ref{fig:cm_cl_all}(c) and \ref{fig:cm_cl_all}(d) show the TEV-induced lift ($C_L^{TEV}$) and moment ($C_M^{TEV}$) for these cases. We again see large variability in the peak loading induced by the TEV. It is particularly interesting to point out the large peak in TEV-induced lift and moment seen in two particular cases, which are indicated using an arrow in figure \ref{fig:cm_cl_all}(c). These cases correspond to low-frequency oscillations which, as seen in the previous section, behave differently from all other cases even in terms of the vortex kinematics. The remainder of this section will focus on demonstrating the efficacy of this framework in analyzing some aspects of the behaviour highlighted here -- particularly the relative importance of the LEV and TEV as the airfoil kinematics are varied, and the phase of their induced loading with respect to the airfoil's motion.
\label{sec:airfoil_vortex_FPM}

\begin{figure}
  \centerline{\includegraphics[scale=1.0]{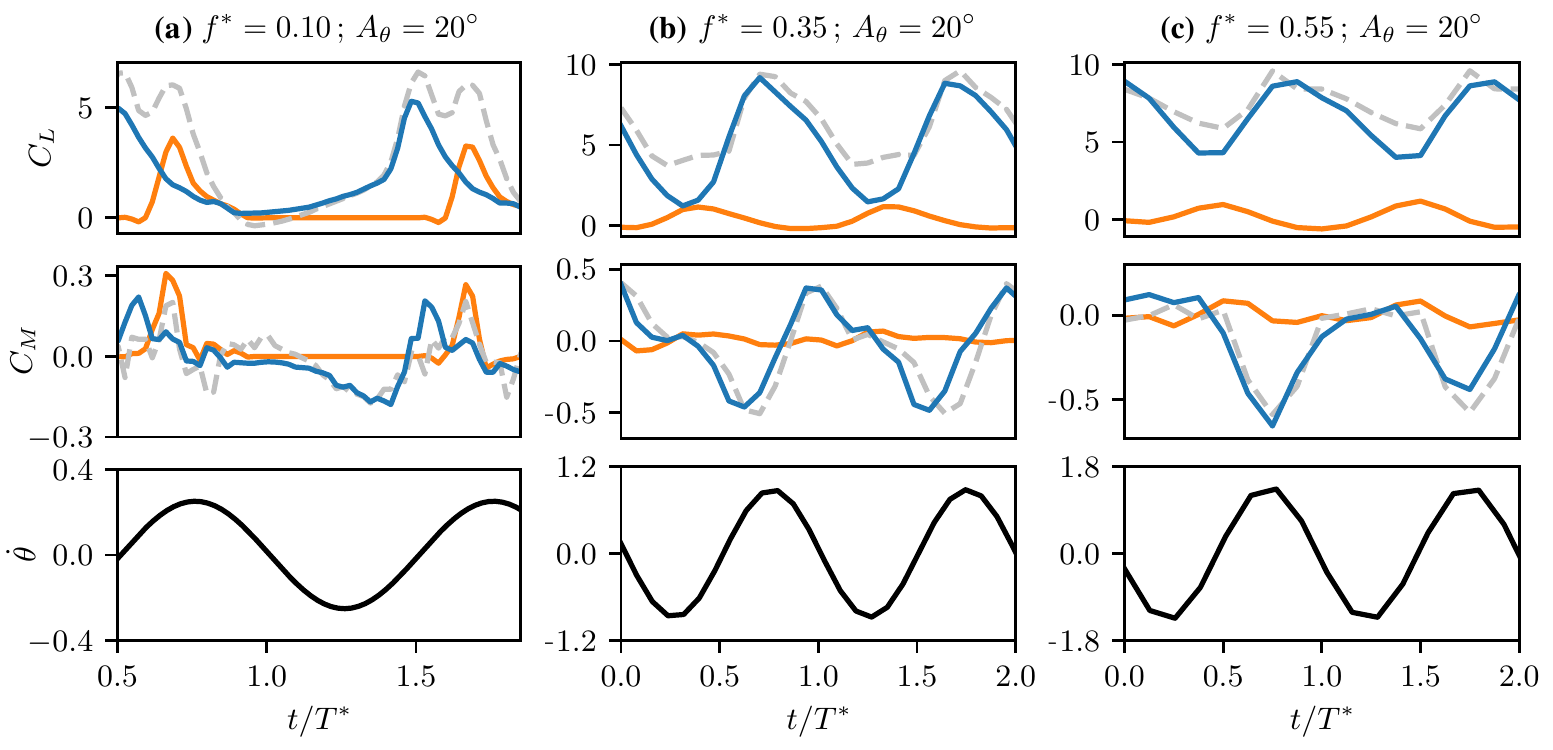}}
  \caption{Top panel shows $C_L$ induced by the LEV (\protect\blueline) and TEV (\protect\orangeline), compared with total $C_L$ induced by all vortical regions (\protect\greydashedline). Middle panel shows $C_M$ induced by the LEV (\protect\blueline) and TEV (\protect\orangeline), compared with total $C_M$ induced by all vortical regions (\protect\greydashedline). Bottom panel shows angular velocity ($\dot{\theta}$). All quantities are plotted against phase of oscillation ($t/T^*$) for three cases representing different vortex dynamic regimes. (a) $f^*=0.10$, $A_{\theta}=20^{\circ}$; (b) $f^*=0.35$, $A_{\theta}=20^{\circ}$; (c) $f^*=0.55$, $A_{\theta}=20^{\circ}$. }
\label{fig:lift_moment}
\end{figure}

We now focus our discussion on some select cases in the data-set, and delve further into the details of the aerodynamic loading induced by the LEV and TEV. In figure \ref{fig:lift_moment}, we compare the vortex-induced force and moment in three cases representing distinct vortex-dynamic regimes, with labels $(5)$, $(2)$ and $(13)$. These cases correspond to airfoils oscillating with the same amplitude and at different frequencies. The frequencies of oscillation are $f^*=0.10$, $f^*=0.35$, and $f^*=0.55$ in figures \ref{fig:lift_moment}(a), \ref{fig:lift_moment}(b) and \ref{fig:lift_moment}(c) respectively. The amplitude of oscillation is $A_{\theta}=20^{\circ}$ for all three cases. The top panel for each case shows the phase-averaged coefficient of lift ($C_L$) induced by the LEV (\blueline; blue solid line), TEV (\orangeline; orange solid line), and the total $C_L$ induced by all vortex structures (\greydashedline; grey dashed line). Similarly, the middle panel shows the coefficient of moment ($C_M$) induced by the LEV, TEV, and all vortex structures. In the bottom panel, the force and moment time-series are compared with the phase of the angular velocity of the airfoil. The oscillation phase is represented as $t/T^*$ on the $X$-axis.

It is immediately apparent on comparing the time-series plots of the $C_L$ and $C_M$ induced by the LEV (\blueline; blue solid line) with the total induced by all vortex structures (\greydashedline; grey dashed line), that the LEV accounts for the bulk of the vortex-induced force as well as moment production. In fact, this is true of nearly all the cases analyzed in this work. Moreover, these cases show that the LEV dictates a larger proportion of the force and moment production as the oscillation frequency is increased. We also see that the aerodynamic loading due to the TEV (\orangeline; orange solid line) is comparable to that of the LEV only in the lowest-frequency case at $f^*=0.10$, and is negligible at higher frequencies. This is related to the delayed separation of the TEV that was observed at low frequencies using the vortex trajectories in figure \ref{fig:clust_traj}. This delayed separation therefore allows the TEV to benefit from the feeding shear-layer on the pressure side of the airfoil for a longer duration. 

Another important aspect of the forces and moments induced by these vortex structures is their phase with respect to the oscillation of the airfoil. 
The phase difference between the force/moment production and the kinematics can be shown to be a primary determinant of the energy extracted by an elastic structure from the surrounding flow \citep{Menon2019,Menon2020DynamicAirfoil}. This energy extraction is in turn responsible for the initiation and sustenance of flow-induced oscillations, and is an important quantity in the study of aeroelastic response branches and their stability \citep{Morse2009,Kumar2016Lock-inCylinder,Menon2019,Menon2020AeroelasticMaps,Zhu2020NonlinearWing,Menon2020OnPartitioning}. In the context of pitching airfoils, the dimensionless energy extracted by the pitch degree-of-freedom over an oscillation period ($T^*=1/f^*$) can be written as shown in equation \ref{eq:energy}. Contours of this extracted energy as a function of kinematics were shown in figure \ref{fig:vortex_intro} for the current data-set (reproduced from ref. \citep{Menon2019}). It can be shown that $E^*$ depends on the phase difference between $C_M$ and $\dot{\theta}$, and is maximum when they are in-phase \citep{Menon2019}. In the present case, we see from figures \ref{fig:lift_moment}(a)-(c) that $\dot{\theta}$ and the LEV-induced $C_M$ get progressively more out-of-phase as the oscillation frequency is increased. In addition, there is also a variation in the phase of $C_L$ with respect to the kinematics of the airfoil. This is relevant in multi-degree-of-freedom applications such as energy harvesting. In particular, for systems based on energy extraction from heave oscillations that are instigated by pitching-related non-linearities \citep{Zhu2009ModelingHarvester}, the phase of the LEV-induced $C_L$ with respect to the oscillation is an important aspect in their performance. This method can therefore inform design considerations in these applications to either enhance or diminish the energy extraction from various flow-structures, as necessary. 

\begin{figure}
  \centerline{\includegraphics[scale=1.0]{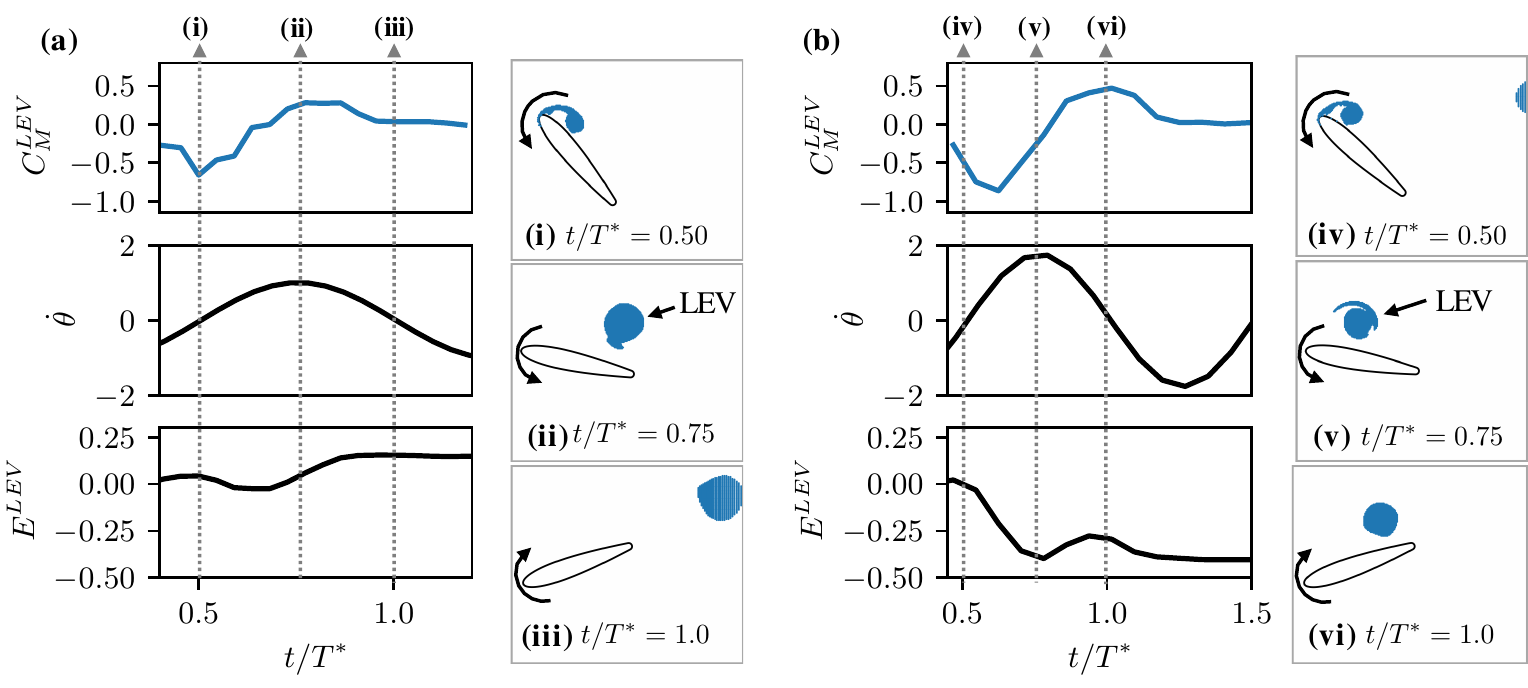}}
  \caption{Comparison of the phase difference between LEV-induced moment and angular velocity for two cases: (a, i--iii) $f^*=0.20$, $A_{\theta}=35^{\circ}$; and (b, iv--vi) $f^*=0.35$, $A_{\theta}=35^{\circ}$. For each case, time-series plots show phase-averaged LEV-induced coefficient of moment (top panel; $C_M^{LEV}$), angular velocity (middle panel; $\dot{\theta}$), and energy extraction from LEV-induced moment (bottom panel; $E^{LEV}$). Snapshots for each case show the LEV's position at three phases of oscillation, $t/T^* = 0.5$, $t/T^* = 0.75$ and $t/T^* = 1.0$ (indicated by vertical dashed lines). Note: Angular velocity and moment are positive in the anti-clockwise (pitch-down) direction.}
\label{fig:phase_snapshots}
\end{figure}

As a final demonstration of the current method, we use flow-field data in conjunction with the computed forces and moments in order to explain the phase-behaviour observed above. In particular, we analyze the physics behind the observation that the LEV-induced moment is more out-of-phase with angular velocity as the oscillation frequency is increased. For this analysis we focus on two cases which have oscillation frequencies $f^* = 0.20$ and $f^* = 0.35$, with equal amplitudes of $A_{\theta}=35^{\circ}$. These cases are chosen because they correspond to positive and negative energy extraction ($E^*$) respectively, or favourable and unfavourable phase difference between $C_M$ and $\dot{\theta}$. In addition, these cases lie very close to the boundary between positive and negative energy extraction by the pitching airfoil (see $E^*=0$ contour in figure  \ref{fig:vortex_intro}), and therefore represent a situation where an increase in oscillation frequency (at constant amplitude) causes $C_M$ to go from in-phase to out-of-phase with respect to $\dot{\theta}$. The cases with $f^* = 0.20$ and $f^* = 0.35$ are labelled $(6)$ and $(10)$ in the preceding discussion, and their analysis is presented in figures \ref{fig:phase_snapshots}(a) and \ref{fig:phase_snapshots}(b) respectively.

In figure, \ref{fig:phase_snapshots} the top panel shows the moment induced by the LEV for each case, denoted as $C^{LEV}_M$. This is compared with $\dot{\theta}$ in the middle panel. We note that moment and angular velocity are positive in the anti-clockwise direction. The bottom panel shows the LEV-induced energy extraction for each case, computed, as a function of time. This is the energy extracted exclusively from the LEV-induced moment, $C^{LEV}_M$, and is given by,
\begin{equation}
    E^{LEV}(t) = \int^{t}_{0} C^{LEV}_M \ \dot{\theta} \ dt.
    \label{eq:lev_energy}
\end{equation}
We first point out that the energy extracted from the LEV is indeed positive for the lower frequency case in figure \ref{fig:phase_snapshots}(a) and negative for the higher frequency case in figure \ref{fig:phase_snapshots}(b). This therefore confirms that there is a phase-shift between the $C^{LEV}_M$ and $\dot{\theta}$ on increasing the oscillation frequency. In particular, it is easy to see that the phase of maximum $\dot{\theta}$ matches well with the $C_M$-maximum in the case of $f^*=0.20$ in figure \ref{fig:phase_snapshots}(a), whereas these maxima are shifted by approximately $90^{\circ}$ from each other for the case of $f^*=0.35$ in figure \ref{fig:phase_snapshots}(b). In order to relate these time-series plots with the dynamics of the flow, figures \ref{fig:phase_snapshots}(i)-(iii) and \ref{fig:phase_snapshots}(iv)-(vi) show instantaneous snapshots of the LEV and airfoil at three phases of oscillation. Here $t/T^*=0.5$ represents the beginning of the pitch-down motion from the maximum angle-of-attack, $t/T^*=0.75$ occurs at the mean angle-of-attack during the pitch-down motion, and $t/T^*=1.0$ is the end of the pitch-down motion. The phase of each snapshot is indicated by the vertical dashed lines in figures \ref{fig:phase_snapshots}(a) and \ref{fig:phase_snapshots}(b).

In both cases, we see from the snapshots in figures \ref{fig:phase_snapshots}(i) and \ref{fig:phase_snapshots}(iv) that the LEV is attached to the airfoil at the start of the pitch-down stroke. As seen in the $C^{LEV}_M$ plots at $t/T^*=0.5$, it induces a strong pitch-up (negative) moment on the airfoil in both cases, due to its proximity to the leading-edge. In the low-frequency case however, $t/T^*=0.5$ is immediately followed by a reduction in the pitch-up (negative) moment. Noting that $t/T^*=0.5$ corresponds to the start of the pitch-down stroke, this indicates that the LEV separates very soon after the airfoil begins pitching-down in the $f^*=0.20$ case. This is in contrast to the higher-frequency case, where we see from figure \ref{fig:phase_snapshots}(b) that the negative peak in $C^{LEV}_M$ occurs later in the cycle (at $t/T^* \approx 0.625$). Although not shown here, it is easy to verify from flow snapshots that the earlier separation of the LEV in the low-frequency case occurs because the slower, low-frequency motion allows the LEV to begin developing earlier in the previous pitch-up stroke. Compared to higher-frequency motion, this early roll-up and slower pitching velocity allows more time for the LEV's growth and saturation, hence promoting earlier shedding.

This phase-difference in the LEV's growth and shedding has significant consequences as the airfoil reaches its maximum pitch-down angular velocity at $t/T^* = 0.75$. For the low-frequency case at $t/T^* = 0.75$, figure \ref{fig:phase_snapshots}(ii) shows that the LEV has convected past the hinge location (mid-chord) and is positioned over the downstream half of the airfoil. In this configuration, the LEV-induced suction generates a peak in pitch-down (positive) moment, which coincides with the peak in angular velocity. This is seen at $t/T^* = 0.75$ in figure \ref{fig:phase_snapshots}(a). This favourable timing between the LEV and airfoil kinematics therefore results in $C^{LEV}_M$ being in-phase with $\dot{\theta}$ in the low-frequency case, and the airfoil extracting positive energy from the LEV. Subsequently, the LEV convects farther downstream, and $C^{LEV}_M$ becomes negligible.

In the high-frequency case, on the other hand, snapshot \ref{fig:phase_snapshots}(v) shows that the LEV has not yet convected past the hinge location at $t/T^* = 0.75$, when the airfoil attains its maximum pitch-down velocity. This is due to the airfoil's faster motion as well as the delayed separation of the LEV. As a consequence, the plot of $C^{LEV}_M$ in figure \ref{fig:phase_snapshots}(b) shows that the LEV-induced moment is small and slightly negative at $t/T^* = 0.75$, when $\dot{\theta}$ is at its peak. In this case, the maximum value in $C^{LEV}_M$ is attained later in the cycle, at $t/T^* = 1.0$. As seen in figure \ref{fig:phase_snapshots}(vi), the LEV is located above the downstream half of the airfoil at $t/T^* = 1.0$, and consequently generates its peak pitch-down moment at this time-instance. However, $t/T^* = 1.0$ corresponds to the start of the pitch-up stroke. Due to the fact that the maximum pitch-down moment is induced by the LEV at the start of the airfoil's pitch-up motion, the phase between $C^{LEV}_M$ and $\dot{\theta}$ is unfavourable and hence leads to negative energy extraction from the LEV.

This analysis highlights the competing influence of the pitching timescale and important flow timescales (such as those for vortex shedding and convection) in determining the phase of the forcing with respect to the kinematics. This also demonstrates that in addition to accurately tracking and quantifying the kinematic and dynamic effects of particular vortex structures, this framework also allows us to precisely correlate these estimated quantities with observed flow phenomena.

\section{Summary}
\label{sec:summary}

In this work, we have presented a data-driven and physics-based computational framework for the analysis of vortex-dominated flows. The main focus is a flexible and automated method to accurately evaluate kinematic quantities and the aerodynamic loading of individual vortex structures in complex  vortex-dominated flows. This method uses a novel force and moment partitioning formulation which breaks down the aerodynamic loading on an immersed body into physically insightful components. In the particular form of the formulation used here, we make a direct connection between vorticity-induced local deformation of the flow-field, via the $Q$-criterion, and its effect on forces and moments induced on an immersed body. The force and moment partitioning method also allows the rigorous evaluation of the loading due to specific vortical regions in the flow-field. This physics-based formulation is combined with a suite of physics-informed and data-driven methods to simultaneously detect, isolate, segment, track, and categorize several distinct vortices in complex flow-fields. The end result is a framework that takes in time-resolved flow fields and provides quantitative details of the kinematic evolution as well as the aerodynamic loading due to each vortex structure in the problem. 


We present an application of these methods to a large data-set of 165 two-dimensional pitching airfoil simulations at a wide range of kinematic operating conditions. We first demonstrate a data-driven rank-reduction procedure to identify unique vortex-dynamic regimes within this data-set, and then deploy the aforementioned analysis framework to analyze this set of distinct regimes. The analysis reveals several interesting aspects of the vortex kinematics, such as period-doubling in vortex trajectories and the dependence of circulation on the airfoil kinematics. These are generally non-trivial to extract from such large ensembles of flow-field data. Further, the utility of this method in analyzing the dynamical influence of key vortex structures is also demonstrated. In particular, we quantify the force and moment induced by leading and trailing-edge vortices for various cases, and highlight how their relative importance varies with the kinematics of the airfoil's oscillation. We also analyze the phase between the forces/moments generated and the motion of the airfoil, and connect this to their relevance in flow-induced oscillation and energy harvesting.

It is important to highlight that while the present analysis focuses exclusively on vortex structures (detected as regions of $Q>0$), the framework is equally applicable to the analysis of other flow structures that are defined by some scalar field. Further, several aspects in the current implementation of the vortex analysis as well as vortex pattern-based rank-reduction methods can be modified due to the flexible nature of the framework. For instance, while vortices are identified using the $Q$-criterion in this implementation, a variety of other tools aimed at detecting coherent structures can be used instead \citep{Haller2015LagrangianStructures,Hadjighasem2017ADetection}. The DBSCAN-based method for isolating and segmenting these vortices can also be replaced by spectral and graph clustering techniques \citep{Hadjighasem2016Spectral-clusteringDetection,Schlueter-Kuck2017CoherentTheory,Nair2015Network-theoreticDynamics}. Additionally, the grouping of these vortices can be performed based on vortex shape, size, as well as dynamical influence, in place of the trajectory and circulation-based grouping used here.  
In terms of limitations, the applicability of these methods to flows with vortex structures that are less coherent, such as in turbulent flows, has not been demonstrated here. Further, we have not demonstrated its use in complex, three-dimensional flow-fields. The extension of this framework to three-dimensional problems is a subject of ongoing work. However, it should be noted that the two-dimensional dynamics of the flow are fundamental to many physically relevant problems, and a two-dimensional analysis such as that presented here is very insightful in these situations.

\section*{Acknowledgments}
This work is supported by the Air Force Office of Scientific Research Grant Number FA 9550-16-1-0404, monitored by Dr. Gregg Abate. This work also benefited from the computational resources at Extreme Science and Engineering Discovery Environment (XSEDE), which is supported by National Science Foundation grant number ACI-1548562, through allocation number TG-CTS100002, and at the Maryland Advanced Research Computing Center (MARCC).

\appendix

\section{Derivation of the moment (and force) partitioning method}
\label{app:fpm_derivation}
\newcommand\numberthis{\addtocounter{equation}{1}\tag{\theequation}}

Here we provide a detailed derivation of the moment partitioning method described in section \ref{sec:fpm}. The force partitioning is derived in a similar manner, and the connection between the two is discussed at the end of this section. Additionally, the reader is referred to ref. \citep{Menon2020OnPartitioning} for a detailed derivation of an alternate form for the force partitioning method. We note that some aspects of the discussion in section \ref{sec:fpm} are repeated here for the sake of a self-contained derivation.

We begin with defining the setup of the moment partitioning problem. We would like to partition the flow-induced moment in the $k$-direction on an immersed body. The surface of the immersed body is denoted as $B$, and it is immersed in a fluid domain where the volume contained by the fluid, external to the body, is denoted as $\vol$. This fluid domain is bounded externally by the surface $\Sigma$, and internally by $B$. The unit vector $\hat{n}$, defining the orientation at every point along the bounding surfaces $B$ and $\Sigma$, points out from the fluid volume (into the surfaces $B$ and $\Sigma$), and the position vector of every point along these surfaces is denoted by $\vec{X}$. The point on the body about which the moment is calculated is given by $\vec{X_c}$.

We start the derivation with the Navier-Stokes momentum equation, written in the following form,
\begin{equation}
  \frac{\partial \vec{u}}{\partial t} + \vec{u} \cdot \vec{\nabla} \vec{u} = - \vec{\nabla} p + \frac{1}{Re} \vec{\nabla}^2 \vec{u} 
  \label{eq:lamb-gromeka}
\end{equation}
An auxiliary potential, $\psi_{k}$, is constructed at every time-instance, which is a function of the instantaneous position and shape of the immersed body as well as the outer domain boundary. This potential is defined as:
\begin{equation}
  \vec{\nabla}^2 \psi_{k} = 0, \ \ \mathrm{ with } \ \
  \hat{n} \cdot \vec{\nabla} \psi_{k}=
    \begin{cases}
     \big[ (\vec{X}-\vec{X_c})\times\hat{n} \big]\cdot \hat{e}_k \;, \; \mathrm{on} \; B \\
      0 \; \;, \; \mathrm{on} \; \Sigma \\
    \end{cases}
  \label{eq:mpm_scalar_deriv}
\end{equation}
The Navier-Stokes equation (\ref{eq:lamb-gromeka}) is now projected on to the gradient of this auxiliary potential, and the result is integrated over the volume of the fluid domain, $\vol$:
\begin{align*}
  \int_\vol \frac{\partial \vec{u}}{\partial t} \cdot \vec{\nabla} \psi_{k} dV + \int_\vol \big( \vec{u} \cdot \vec{\nabla} \vec{u} \big) \cdot \vec{\nabla} \psi_{k} dV 
  = -\int_\vol \vec{\nabla} p \cdot \vec{\nabla} \psi_{k} dV + \frac{1}{Re} \int_\vol \big( \vec{\nabla}^2 \vec{u} \; \big) \cdot \vec{\nabla} \psi_{k} dV \numberthis
  \label{eq:projection}
\end{align*}

We now simplify each term of the above equation separately, starting with the pressure term (first term) on the right-hand side.
\begin{equation}
   \int_\vol \vec{\nabla} p \cdot \vec{\nabla} \psi_{k} dV  = \int_\vol \vec{\nabla} \cdot (p \vec{\nabla} \psi_{k}) dV = \int_{B+\Sigma} p \hat{n} \cdot \vec{\nabla} \psi_{k} dS = \int_{B} p \big[ (\vec{X}-\vec{X_c})\times\hat{n} \big]\cdot \hat{e}_k dS
  \label{eq:pressure_final}
\end{equation}
where we use the divergence theorem and the last step follows from the boundary condition on the field $\psi_{k}$. Equation \ref{eq:pressure_final} is evidently the moment induced on the body due to the surface pressure distribution exerted by the surrounding fluid.

The viscous term, which is the second term on the right-hand side of equation \ref{eq:projection}, is treated as follows:
\begin{align*}
  \int_\vol \big( \vec{\nabla}^2 \vec{u} \; \big) \cdot \vec{\nabla} \psi_{k} dV = -\int_\vol (\vec{\nabla} \times \vec{\omega}) \cdot \vec{\nabla} \psi_{k} dV 
  = -\int_{B+\Sigma} \hat{n} \cdot ( \vec{\omega} \times \vec{\nabla} \psi_{k}) dS = \int_{B+\Sigma} (\vec{\omega} \times \hat{n} ) \cdot \vec{\nabla} \psi_{k} dS \numberthis
  \label{eq:viscous}
\end{align*}

Finally, we simplify the unsteady term (first term in equation \ref{eq:projection}) as follows:
\begin{equation}
  \int_\vol \frac{\partial \vec{u}}{\partial t} \cdot \vec{\nabla} \psi_{k} dV = \int_\vol \vec{\nabla} \cdot \bigg( \frac{\partial \vec{u}}{\partial t} \psi_{k} \bigg) dV = \int_{B+\Sigma} \hat{n} \cdot \bigg( \frac{d \vec{u}}{dt} \psi_{k} \bigg) dS - \int_\vol \vec{\nabla} \cdot \Bigg [ \Big( \vec{u} \cdot \vec{\nabla} \vec{u} \Big)\psi_{k} \Bigg] dV
  \label{eq:unsteady_final}
\end{equation}
where the first step in the above expression uses the incompressibility constraint, $\vec{\nabla} \cdot \vec{u}=0$, and the second step follows from the divergence theorem and the definition of the Lagrangian derivative, $d \vec{u} /dt = \partial \vec{u}/\partial t + \vec{u} \cdot \vec{\nabla} \vec{u}$. 

We now plug equations \ref{eq:pressure_final}, \ref{eq:viscous} and \ref{eq:unsteady_final} into equation \ref{eq:projection}. After rearranging and simplifying terms we have,
\begin{align*}
  \int_{B} p \big[ (\vec{X}-\vec{X_c})\times\hat{n} \big]\cdot \hat{e}_k dS = &- \int_{B+\Sigma} \hat{n} \cdot \bigg( \frac{d \vec{u}}{dt} \psi_{k} \bigg) dS + \int_\vol \Big[ \vec{\nabla} \cdot \Big( \vec{u} \cdot \vec{\nabla} \vec{u} \Big) \Big] \psi_{k} dV + \int_{B+\Sigma} \frac{1}{Re} (\vec{\omega} \times \hat{n}) \cdot \vec{\nabla} \psi_{k} dS \numberthis
  \label{eq:pre-final}
\end{align*}
where the three terms on the right-hand side correspond to the pressure loading due to unsteady effects, fluid velocity gradients in the flow around the immersed body, and viscous effects on the surfaces respectively. 

We can derive further physical insight into the flow gradient-related effects, i.e. the second term on the right-hand side of \ref{eq:pre-final}, by relating this term to local flow kinematics described by the velocity-gradient tensor, $\vec{\nabla}\vec{u}$. In particular, it can be shown that the second-invariant of $\vec{\nabla}\vec{u}$, which is referred to as $Q$, can be directly related to this term in the following manner:
\begin{align}
Q = \frac{1}{2}\bigg[ \mathrm{Tr}\Big(\vec{\nabla}\vec{u} \Big)^2 - \mathrm{Tr}\Big(\vec{\nabla}\vec{u}^2 \Big) \bigg] = -\frac{1}{2}\vec{\nabla} \cdot \Big( \vec{u} \cdot \vec{\nabla} \vec{u} \Big)
\end{align}
In the above expression, $\mathrm{Tr}(\cdot)$ is the trace of a tensor. We note here that $Q = (1/2)\big(||\boldsymbol{\Omega}||^2 - ||\boldsymbol{S}||^2 \big)$ also corresponds to a comparison of local rotation versus strain in the flow-field, where $\boldsymbol{\Omega}$ and $\boldsymbol{S}$ are the anti-symmetric and symmetric parts of $\vec{\nabla}\vec{u}$ respectively. 
This result can be used in equation \ref{eq:pre-final}, to rewrite the decomposition of pressure-induced moments as follows:
\begin{align*}
  \int_{B} p \big[ (\vec{X}-\vec{X_c})\times\hat{n} \big]\cdot \hat{e}_k dS = - \int_{B+\Sigma} \hat{n} \cdot \bigg( \frac{d \vec{u}}{dt} \psi_{k} \bigg) dS - \int_\vol 2Q \; \psi_{k} dV 
  + \int_{B+\Sigma} \frac{1}{Re} (\vec{\omega} \times \hat{n}) \cdot \vec{\nabla} \psi_{k} dS \numberthis
  \label{eq:pre-final_q}
\end{align*}
It is now clear from the above equation that the second term on the right-hand side directly relates local flow deformation to the generation of moments on the immersed body. In particular, the sign of $Q$ allows us to separate rotation-induced loading from strain-induced loading.

Additional insight into the physical relevance of the terms in equation \ref{eq:pre-final_q} can be gained by separating the contributions of rotational (vortical) from irrotational flow components in the force/moment production. This is done using the Helmholtz decomposition \citep{BATCHELOR} as follows:
\begin{equation}
  \vec{u} = \vec{u}_\Phi + \vec{u}_v = \vec{\nabla} \Phi + \vec{\nabla} \times A
  \label{eq:helmholtz_deriv}
\end{equation}
where $\Phi$ and $A$ are scalar and vector potentials respectively. The velocity components $\vec{u}_\Phi$ and $\vec{u}_v$ in the above decomposition are the irrotational and rotational (vortical) components of the velocity field respectively. 

Using this decomposition, we can write $Q = Q_{\Phi} + Q_v$. Here $Q_{\Phi} = -\vec{\nabla} \cdot \big( \vec{u}_\Phi \cdot \vec{\nabla} \vec{u}_\Phi \big)/2$ is purely irrotational, and $Q_v = -\big[\vec{\nabla} \cdot \big( \vec{u}_v \cdot \vec{\nabla} \vec{u}_v \big) + \vec{\nabla} \cdot \big( \vec{u}_v \cdot \vec{\nabla} \vec{u}_\Phi \big) + \vec{\nabla} \cdot \big( \vec{u}_\Phi \cdot \vec{\nabla} \vec{u}_v \big)\big]/2$ is non-zero only in the presence of non-zero rotational flow, i.e. where $\vec{u}_v \neq 0$ (which implies $\vec{\omega} \neq 0$). The second term in the right-hand side of equation \ref{eq:pre-final_q} can then be decomposed into contributions from vortical and irrotational effects in the following manner:
\begin{align*}
  -\int_\vol 2Q \; \psi_{k} dV = -\int_\vol 2Q_v \; \psi_{k} dV -\int_\vol 2Q_{\Phi} \; \psi_{k} dV &= -\int_\vol 2Q_v \; \psi_{k} dV + \int_\vol \vec{\nabla} \cdot \Big( \vec{u}_{\Phi} \cdot \vec{\nabla} \vec{u}_{\Phi} \Big) \; \psi_{k} dV \\ &= -\int_\vol 2Q_v \; \psi_{k} dV + \int_\vol \vec{\nabla} \cdot \Big[ \vec{\nabla} \Big( \frac{1}{2}\vec{u}_{\Phi} \cdot \vec{u}_{\Phi} \Big) \Big] \; \psi_{k} dV \numberthis
  \label{eq:helmholtz_q}
\end{align*}
In the above equation, each term on the right-hand side (\ref{eq:helmholtz_q}) has a clear physical significance. The purely irrotational term (last term in the right-hand side of \ref{eq:helmholtz_q}) quantifies the moment generated exclusively by irrotational or potential flow mechanisms in the flow-field. The first term on the right-hand side of \ref{eq:helmholtz_q}, due to its dependence on $Q_v$, is non-zero only in the presence of non-zero vorticity (where $\vec{u}_v \neq 0$). This term therefore corresponds to the component of moment generated by vorticity-induced effects in the flow-field. More specifically, its dependence on $Q_v$ suggests that this is the pressure-moment generated by vorticity-induced strain and rotation in the flow-field.

It can be shown, by following along the lines of ref. \citep{Zhang2015MechanismsInsects}, that the irrotational term in \ref{eq:helmholtz_q} is small in most cases and goes to zero in sufficiently large domains. Therefore for clarity, we refer to this purely irrotational contribution as $\int_\vol \vec{\nabla} \cdot \Big[ \vec{\nabla} \Big( \frac{1}{2}\vec{u}_{\Phi} \cdot \vec{u}_{\Phi} \Big) \Big] \; \psi_{k} dV =\epsilon_{M_k}^{\Phi}$ (noting that $\epsilon_{M_k}^{\Phi} \approx 0$ in most situations), and rewrite the vorticity-induced moment in the following manner:
\begin{align*}
  -\int_\vol 2Q_v \; \psi_{k} dV = -\int_\vol 2Q \; \psi_{k} dV - \epsilon_{M_k}^{\Phi} \numberthis
  \label{eq:vif_q}
\end{align*}

Lastly, we note that equation \ref{eq:pre-final_q} represents the partitioning of the moments induced solely by the pressure distribution on the immersed body. However, the total moment also includes shear contributions. The force coefficient in the $k$-direction due to viscous shear can be written as $C_{F_k}^{\nu} = -(1/Re) \int_B (\vec{\omega} \times \hat{n}) \cdot \hat{e}_k dS$, and therefore the moment induced by this surface shear is given by  $C_{M_k}^{\nu} = -(1/Re) \int_B [(\vec{X}-\vec{X}_c)\times(\vec{\omega} \times \hat{n})] \cdot \hat{e}_k dS$.

Using equations \ref{eq:helmholtz_q}, \ref{eq:vif_q} and the above expression for the moment induced by viscous shear on the surface in equation \ref{eq:pre-final_q}, we arrive at the final form for the partitioning of the total moment on the body:
\begin{align}
  C^{(k)}_M &= - \int_{B} \hat{n} \cdot \bigg( \frac{d \vec{U}_{B}}{dt} \psi_{k} \bigg) dS  \label{eq:kin_deriv} \\
  &- \int_\vol 2Q \; \psi_{k} dV - \epsilon_{M_k}^{\Phi} \label{eq:vif_deriv} \\
  &+ \int_{B} \frac{1}{Re} \bigg\{ (\vec{\omega} \times \hat{n}) \cdot \vec{\nabla} \psi_{k} - (\vec{\omega} \times \hat{n}) \cdot \Big[\hat{e}_k \times (\vec{X}-\vec{X}_c) \Big] \bigg\} dS \label{eq:shr_deriv} \\
  &+ \int_\vol \vec{\nabla} \cdot \Big[ \vec{\nabla} \Big( \frac{1}{2}\vec{u}_{\Phi} \cdot \vec{u}_{\Phi} \Big) \Big] \; \psi_{k} dV \label{eq:pot_deriv} \\
  &+ \int_{\Sigma} \Bigg[ -\hat{n} \cdot \bigg( \frac{d \vec{u}}{dt} \psi_{k} \bigg) + \frac{1}{Re} (\vec{\omega} \times \hat{n}) \cdot \vec{\nabla} \psi_{k} \Bigg] dS \label{eq:ext_deriv}
\end{align}
The expressions in \ref{eq:kin_deriv}, \ref{eq:vif_deriv}, \ref{eq:shr_deriv}, \ref{eq:pot_deriv} and \ref{eq:ext_deriv} correspond to the moment partitioning components $\mpmkin$, $\mpmvif$, $\mpmshr$, $\mpmpot$ and $\mpmext$ respectively shown in equation \ref{eq:fpm_mpm_decomp}. Note that in the above equation, $\epsilon_{M_k}^{\Phi} = \textrm{\ref{eq:pot_deriv}} \approx 0$ in most cases.

The partitioning of forces in the $i$-direction follows along the same lines as the above derivation. The key difference in that case is the definition of the auxiliary potential, $\phi_i$, which is given in equation \ref{eq:fpm_scalar} of section \ref{sec:fpm}. Further, the viscous shear force takes a different form from that shown in \ref{eq:shr_deriv}. As mentioned above, the viscous shear force in the $i$-direction can be written as $C_{F_i}^{\nu} = -(1/Re) \int_B (\vec{\omega} \times \hat{n}) \cdot \hat{e}_i dS$. The final form of the force partitioning is therefore, 
\begin{align}
  C^{(i)}_F &= - \int_{B} \hat{n} \cdot \bigg( \frac{d \vec{U}_{B}}{dt} \phi_{i} \bigg) dS  \label{eq:kin_deriv_fpm} \\
  &- \int_\vol 2Q \; \phi_{i} dV - \epsilon_{F_i}^{\Phi} \label{eq:vif_deriv_fpm} \\
  &+ \int_{B} \frac{1}{Re} \bigg\{ (\vec{\omega} \times \hat{n}) \cdot \vec{\nabla} \phi_{i} - (\vec{\omega} \times \hat{n}) \cdot \hat{e}_i \bigg\} dS \label{eq:shr_deriv_fpm} \\
  &+ \int_\vol \vec{\nabla} \cdot \Big[ \vec{\nabla} \Big( \frac{1}{2}\vec{u}_{\Phi} \cdot \vec{u}_{\Phi} \Big) \Big] \; \phi_{i} dV \label{eq:pot_deriv_fpm} \\
  &+ \int_{\Sigma} \Bigg[ -\hat{n} \cdot \bigg( \frac{d \vec{u}}{dt} \phi_{i} \bigg) + \frac{1}{Re} (\vec{\omega} \times \hat{n}) \cdot \vec{\nabla} \phi_{i} \Bigg] dS \label{eq:ext_deriv_fpm}
\end{align}
The expressions in \ref{eq:kin_deriv_fpm}, \ref{eq:vif_deriv_fpm}, \ref{eq:shr_deriv_fpm}, \ref{eq:pot_deriv_fpm} and \ref{eq:ext_deriv_fpm} correspond to the force partitioning components $\fpmkin$, $\fpmvif$, $\fpmshr$, $\fpmpot$ and $\fpmext$ respectively shown in equation \ref{eq:fpm_mpm_decomp}. As in the moment partitioning, $\epsilon_{F_i}^{\Phi} = \textrm{\ref{eq:pot_deriv_fpm}} \approx 0$ in most situations. 

We see that the terms \ref{eq:kin_deriv} and \ref{eq:kin_deriv_fpm} depends only on the geometry and velocity of the immersed surface $B$, due to which these terms represent the kinematic components of the total force. Further, it can be shown using the Helmholtz decomposition that this term encapsulates the loading generated by the inviscid (potential flow) added-mass and centripetal added-mass respectively. The vorticity-induced components, in \ref{eq:vif_deriv} and \ref{eq:vif_deriv_fpm}, quantifies the moment generated by vorticity-induced deformations in the flow-field. In particular, the sign of $Q$ can be used to evaluate the aerodynamic loading due to vorticity-induced strain-induced ($Q<0$) and rotation ($Q>0$) in the flow-field. The effect of viscous forces and moments on the surface are encapsulated in \ref{eq:shr_deriv} and \ref{eq:shr_deriv_fpm}. Irrotational or potential flow effects, purely due to the irrotational component of the velocity field ($\vec{u}_{\Phi}$) are isolated in \ref{eq:pot_deriv} and \ref{eq:pot_deriv_fpm}. Finally, \ref{eq:ext_deriv} and \ref{eq:ext_deriv_fpm} represent the effect of the velocity field on the outer domain. It can be shown that this term goes to zero for sufficiently large domains \citep{Zhang2015MechanismsInsects}. For further details about these terms, the reader is referred to reference \citep{Zhang2015MechanismsInsects}.

\section{Methodological details relevant to the data-driven analysis of pitching airfoils}
\label{app:supplementary_results}
\begin{figure}
  \centerline{\includegraphics[scale=1.0]{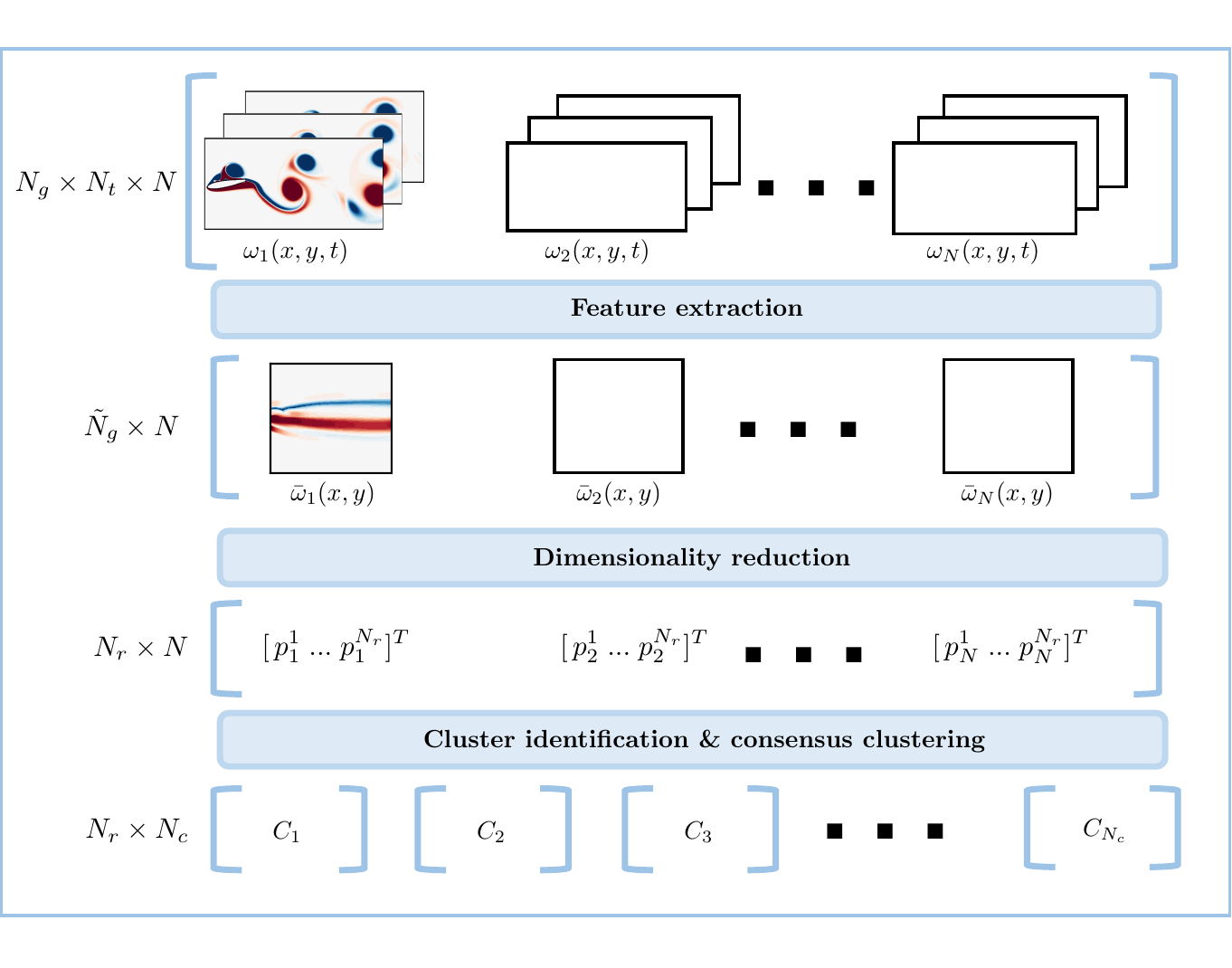}}
  \caption{Schematic of steps involved in the clustering-based framework for the detection of distinct vortex-dynamic regimes. The size of the data at each step is indicated along the left. Snapshots of the vorticity field and the time-averaged vorticity field in the wake for a sample case are shown in $\omega_1$ and $\gamma_1$.}
\label{fig:clustering_schematic}
\end{figure}
In this section, we first describe the specific implementation of the procedure outlined in section \ref{sec:airfoil_wake}, in the context of the identification of distinct vortex dynamic regimes in the wake of pitching airfoils. An outline of the main steps involved in this process is shown in figure \ref{fig:clustering_schematic}.

The first step in identifying distinct vortex shedding regimes in this data-set is the extraction of static ``feature vectors'' representing each of the $N$ cases in the data-set. This is to avoid the complexity associated with discovering similarities in time-resolved data. Here we use the time-averaged vorticity field ($\bar{\omega}$) as the feature vector for each simulation. We use $N_t=500$ time-snapshots of the flow-field for this analysis, and for a grid-size of $N_g$, the initial size of the data set can be estimated as $N_g \times N_t \times N \approx 123 \times 10^3 \times 500 \times 165 \approx 10^{10}$ floating point entries. The extracted feature vector is computed over a region smaller than the total grid-size, and is a $\Tilde{N}_g$-dimensional vector with $\Tilde{N}_g=49392$ grid points. 

Upon identifying static features of the flow for each case in the data-set, we then reduce the dimension of these feature vectors using Principle Component Analysis. To achieve this, we stack the $N$ feature vectors into a $\Tilde{N}_g \times N$ matrix, and compute the principle components of this matrix. The projection of each feature vector on the first $N_r$ principle components then yields a low-dimensional representation of each feature vector. Here $N_r << \Tilde{N}_g$ is chosen so as to retain $\approx 90\%$ of the variance, and we find the required number of modes to be $N_r=25$. The total size of the data-set now comes down to $N_r \times N = 25 \times 165$ floating-point entries, representing a reduction of 6 orders of magnitude from the initial input.

\begin{figure}
  \centerline{\includegraphics[scale=1.0]{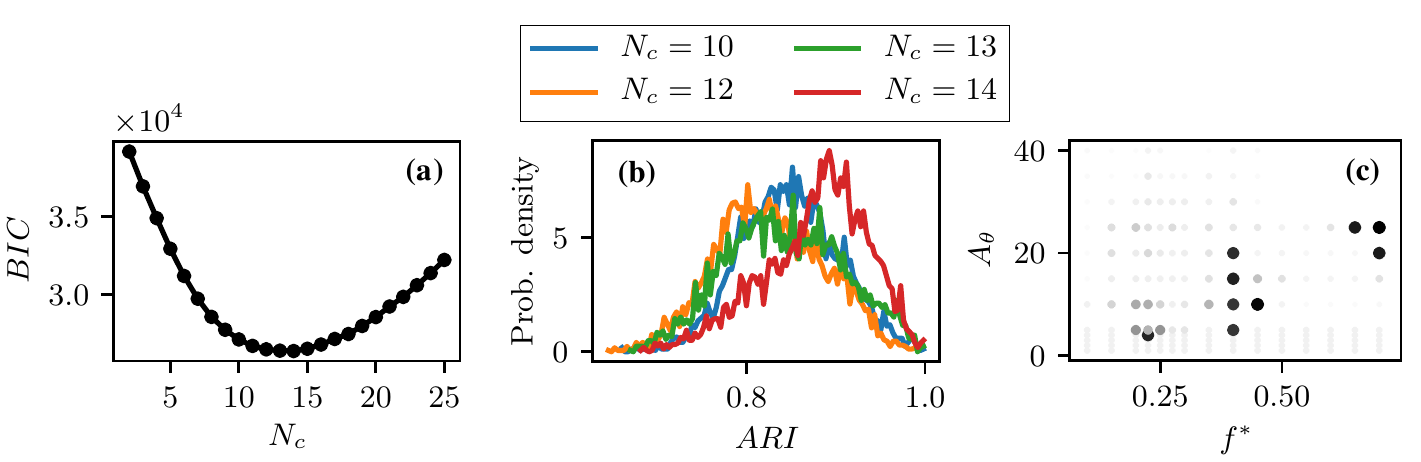}}
  \caption{(a) Bayesian information criterion ($BIC$) versus number of clusters ($N_c$). The minimum $BIC$ at $N_c=14$ indicates the best-fit for $N_c$; (b) Probability density of pair-wise $ARI$ for 100 independent clustering results with $N_c=10$, $N_c=12$, $N_c=13$, and $N_c=14$; (c) Variability in clustering assignment for each case in the dataset, evaluated. The darker markers represent higher variability. }
\label{fig:bic_ari}
\end{figure}
The resulting reduced-dimension data-set is now grouped into clusters that share similar vortex patterns. For this we use Gaussian mixture modelling (GMM), which fits multivariate Gaussian mixtures over given data, where each component Gaussian distribution in the model represents a cluster of similar data \cite{Banfield1993Model-BasedClustering,Fraley1998HowAnalysis,Fraley2002Model-basedEstimation}. This is a simple technique that is also capable of fitting non-isotropic clusters. We fit the model using the expectation-maximization algorithm \citep{Dempster1977MaximumAlgorithm} for maximum-likelihood parameter estimation, implemented in the open-source \textit{scikit-learn} package \citep{Pedregosa2011Scikit-learn:Python}. The number of clusters ($N_c$) in the data is estimated using the Bayesian information criterion \citep{Schwarz1978EstimatingModel}, which penalizes high model complexity and is well suited to mixture modeling based methods \citep{mclachlan2004finite}. The number of clusters is determined by iterating through values in the range $2 \leq N_c \leq 25$, and performing 3000 independent instances of clustering at each $N_c$. We then pick the clustering result that represents the best $BIC$ (from the 3000 independent results) at each value of $N_c$, and this $BIC$ is plotted against $N_c$ in figure \ref{fig:bic_ari}(a). The optimal $N_c$, given by the minimum of the $BIC$-curve, is hence determined to be $N_c=14$ in the data-set analyzed here. 


Lastly, the distinct vortex regimes identified are tested for robustness using multiple independent runs of the GMM algorithm, with random initial seeds, on the given data. In this work, we generate a set of 100 independent clustering results, each of which is selected as the best result (based on $BIC$) from 3000 initializations. We compare independent clustering results pair-wise using the the adjusted Rand index \citep{Hubert1985ComparingPartitions}. This metric assigns a score of $0$ in the event of random clustering, and $1$ in the event of an exact match between two clustering results. We first use this to verify the suitability of the chosen $N_c$ value, with the aim to obtain the smallest possible set of representative cases from this large data-set of flow-fields. Hence we compare the chosen value of $N_c=14$ to clustering results obtained by using smaller $N_c$ values, $N_c = 10$, $N_c = 12$ and $N_c = 13$. We generate a set of 100 independent clustering results at each value of $N_c$ and then compute the $ARI$ between all $99 \times 100$ pairs of results in each set. This is plotted as a probability distribution in figure \ref{fig:bic_ari}(b) for each $N_c$. We see that $ARI \gtrapprox 0.7$ for all pairs, which is a quantitative indication that the data-set consists of highly repeatable clusters, hence suggesting the existence of true clusters (as opposed to random clustering, which would not be repeatable). Further, while the mean $ARI$ for the cases with $N_c=10$, $N_c=12$, and $N_c=13$ are similar to each other, $N_c=14$ has a higher mean $ARI$ as well as a bias towards higher $ARI$ values. This suggests that the chosen value of $N_c=14$ is indeed a better fit to the data than the lower values. This process also allows a quantitative way to determine a ``consensus clustering'' \cite{Monti2003ConsensusData,Vinh2009AClustering}, i.e. the clustering that is most similar to all others generated in this set of independent results. This is chosen based on the mean $ARI$ for each independent clustering, i.e. the mean of its $ARI$ computed against the $99$ other independent results. The clustering result with the lowest mean $ARI$, representing the closest match to all other members of the set, is chosen as the final clustering result \cite{Monti2003ConsensusData,Vinh2009AClustering}.

The repeatability of clustering that is discussed above is assessed over the entire clustering result. We now propose an additional metric for evaluating the repeatability of cluster assignments for each case in the data-set. We construct ``distance matrices'', $M_{ij}$, for each pair ($C_i,C_j$) of independent clustering results, with size $N \times N$ ($N$ is the number of cases in the data-set). The $(p,q)$ entry of $M_{ij}$ is given by the following: $M_{ij}(p,q)=0$ if the $p^{th}$ and $q^{th}$ data-points in the ensemble are clustered either in the same cluster or different clusters in both $C_i$ and $C_j$, and $M_{ij}(p,q)=1$ if the $p^{th}$ and $q^{th}$ points are clustered separately in $C_i$ but together in $C_j$ or vice-versa. This is similar to \citep{Rand1971ObjectiveMethods} and \cite{Gionis2007ClusteringAggregation}. $M_{ij}$ is then averaged for all pairs $(C_i,C_j)$, and subsequently averaged for each data-point $N$. The result is a vector of size $N$ which measures the variability in cluster assignment for each of the $N$ members in the ensemble, thus indicating which ensemble members are most likely to be erroneously assigned to the wrong cluster. In figure \ref{fig:bic_ari}(c), each ensemble member is plotted in frequency-amplitude space, with the visibility of each marker representing the measured variability. As expected, we see that members closer to cluster boundaries are most likely to be mis-classified.

\begin{figure}
  \centerline{\includegraphics[scale=1.0]{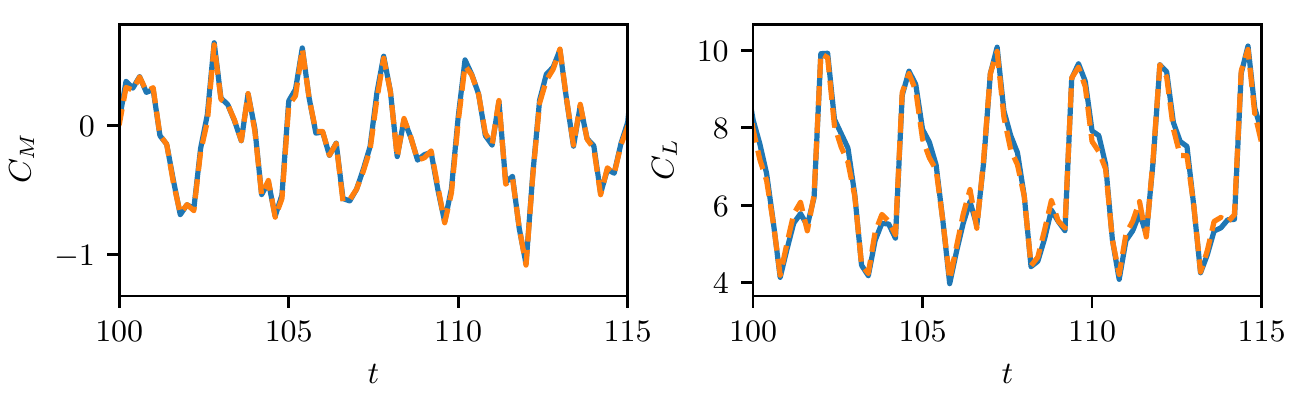}}
  \caption{Comparison of total $C_M$ and $C_L$ induced by all vortex structures detected using the $Q>0$ threshold (\protect\orangedashedline) with those detected using the $Q>5.0$  threshold (\protect\blueline). }
\label{fig:expanded_clust_0.35_20}
\end{figure}
The last aspect of the supplementary results discussed here relates to the vortex tracking and dynamical analysis procedure applied to the pitching airfoils problem. In section \ref{sec:vortex_isolation} we mention that the threshold for vortex detection based on the $Q$-criterion is set as $Q>5.0$ in this work. However, the $Q$-criterion defines vortical regions as those with $Q>0.0$. Therefore we demonstrate here that the threshold of $Q>5.0$ has a minimal influence on the computed forces and moments induced by the detected vortex structures. In figure \ref{fig:expanded_clust_0.35_20} we compare the total $C_L$ and $C_M$ from all vortex structures detected using the $Q>5.0$ threshold (blue, solid line) with those detected using the $Q>0.0$ threshold (orange, dashed line). This corresponds to a representative case in the middle of the data set, with $f^*=0.35$ and $A_{\theta}=20^{\circ}$. We see that the forces agree very well, thus justifying our choice of threshold. 

\bibliographystyle{elsarticle-num} 
\bibliography{references}

\end{document}